\documentclass[11pt,a4paper]{article}

\usepackage[utf8]{inputenc}
\usepackage[T1]{fontenc}
\usepackage[english]{babel}

\usepackage{amsmath}
\usepackage{amsfonts}
\usepackage{amssymb}

\usepackage{graphicx}
\usepackage{booktabs}
\usepackage{array}
\usepackage{multirow}
\usepackage{makecell}
\usepackage{caption}
\usepackage{subcaption}
\usepackage{float}
\usepackage[table,xcdraw]{xcolor}
\usepackage{pdflscape}
\usepackage{threeparttable}

\usepackage[left=2.0cm,right=2.0cm,top=2.0cm,bottom=2.0cm]{geometry}

\usepackage{authblk}

\usepackage{natbib}
\usepackage{breakcites}
\usepackage[bookmarks=false,hidelinks]{hyperref}

\hypersetup{
    colorlinks=true,
    allcolors=blue
}

\usepackage{textcomp}
\usepackage{enumitem}
\usepackage{changepage}
\usepackage{multicol}
\usepackage{soul}
\usepackage[most]{tcolorbox}
\usepackage{setspace}

\usepackage{fancyhdr}
\usepackage[useregional]{datetime2}

\usepackage{sectsty}
\sectionfont{\large}
\subsectionfont{\large\normalfont\itshape}
\subsubsectionfont{\normalfont\itshape}

\title{\vspace{-1cm}\textbf{A Non-Isothermal Viscoplastic Constitutive Model for Clay Slip Surfaces}\vspace{0.4cm}}

\author[1]{Saeed Tourchi\footnote{Corresponding author: \href{mailto:saeed.tourchi@uni.lu}{saeed.tourchi@uni.lu}}}
\author[2,3]{Ehsan Badakhshan}
\author[1]{Milad Jabbarzadeh}
\author[1]{Arash A. Lavasan}
\author[2,3]{Jean Vaunat\vspace{0.4cm}}

\affil[1]{Department of Engineering, University of Luxembourg, Luxembourg}
\affil[2]{Department of Civil and Environmental Engineering, Universitat Politecnica de Catalunya (UPC), Barcelona, Spain}
\affil[3]{International Centre for Numerical Methods in Engineering (CIMNE), Barcelona, Spain}

\date{}

\begin{document}

\begingroup
\setstretch{1.05}
\setlength{\parskip}{0.25em}
\setlength{\parindent}{1.5em}

\maketitle
\vspace{-1.4em}

\begin{abstract}
\small
\setstretch{1.05}
\noindent Clayey slip surfaces play a central role in the reactivation and long-term deformation of slow-moving landslides, particularly when they are subjected to thermal fluctuations from climate, seasonal ground-temperature changes, or subsurface heat sources. Although experimental evidence shows that residual shear strength is sensitive to both temperature and shearing rate, most numerical approaches still treat slip surfaces using temperature-independent strength parameters. This paper presents a non-isothermal viscoplastic constitutive model for clayey slip surfaces implemented through zero-thickness interface elements. The model accounts for temperature-dependent normal and tangential stiffness, progressive degradation of cohesion and friction angle, and rate-dependent viscoplastic slip using a non-associated flow rule. The formulation is coupled with hydraulic and thermal balance equations, allowing the interface response to evolve with stress state, temperature, aperture, and accumulated irreversible displacement. The model is validated against temperature-controlled drained ring-shear tests on bentonite and smectite-rich soils under heating--cooling, cooling--heating, and combined thermal paths. The simulations reproduce the observed rate-dependent response, including thermal strengthening at slow shearing rates and thermal weakening or limited thermal sensitivity at higher rates. The model is further applied to the Congress Street cut benchmark to investigate the effect of thermal cycles on slope stability. The results show that zero-thickness elements improve the representation of strain localization and progressive failure, while increasing temperature progressively degrades interface strength, increases displacement, joint aperture, shear strain, and accelerates joint sliding. The findings demonstrate that temperature-dependent interface degradation can reduce the apparent stability margin of clayey slopes and should be considered in slope stability assessments involving thermal fluctuations.

\vspace{0.4em}

\end{abstract}

\endgroup

\setstretch{1.2}


\section{Introduction}
\label{sec_introduction}
Landslides represent a major geotechnical hazard, causing substantial socio-economic and infrastructural damage, particularly in clay-rich formations where slip surfaces frequently develop along pre-existing shear bands \citep{Petley2012, Hungr2014, Vaunat2002, Alonso1990, Gens2001}. These failures are often long-term processes triggered by cumulative effects of mechanical stress, changes in pore water pressure, and environmental drivers such as rainfall, excavation, or thermal loading. Traditional slope stability analyses typically adopt simplified limit equilibrium approaches with peak or residual shear strength criteria, often neglecting the influence of coupled thermo-hydro-mechanical (THM) processes that govern long-term deformations \citep{Cui2000, Romero2001, Delage2000, Tang1988, Alonso2005}. In expansive or highly plastic clays such as bentonite or illite-bearing shales, temperature variations notably affect the stiffness, permeability, and residual strength of the material \citep{Laloui2003, Hueckel1990, Gens2009}. These effects become particularly critical in reactivated landslides, where slip surfaces are often localized within stiff, low-porosity clays or clay shales. In such geomaterials, coupled thermo-hydro-mechanical interactions can lead to a buildup of excess pore pressure under heating, thereby reducing effective stress and potentially triggering delayed slope reactivation or failure. This behavior is especially relevant in slopes subjected to thermal loads from climatic variations, wildfires, or anthropogenic sources such as shallow geothermal activity \citep{Sheikhahmadi2024, Barla2023, Garakani2022, Villar2008}.

Experimental investigations have consistently revealed that residual shear strength, defined as the minimum resistance mobilized after large displacements, is highly sensitive to both temperature and strain rate \citep{Skempton1985, Tika1999, Tika1996, Stark1994, Lupini1981}. Using ring-shear and torsional shear devices, researchers have shown that heating clay rich soils leads to a marked reduction in residual friction angle and cohesion, attributed to bond softening, suction dissipation, and structural realignment at the microscale \citep{Tang1988, Delage2000, Romero2011, Alonso2020, Shibasaki2017}.The response of clay-rich slip surfaces to thermal loading is highly dependent on both shear rate and material composition. Under slow shearing, particularly in compacted or partially saturated clays with high smectite content, heating can enhance residual strength due to increased suction and microstructural realignment \citep{Shibasaki2017, Loche2023}. In contrast, when shearing occurs at higher rates, the suppression of pore pressure dissipation can lead to thermal weakening, characterized by a reduction in effective stress and residual friction angle \citep{Zhang2021, Barla2023}. This dual behavior, thermal strengthening at low shear rates and thermal weakening at high shear rates, has been consistently observed in laboratory studies on compacted bentonites, reactivated stiff clays, and natural slip zones subjected to controlled thermal and mechanical loads \citep{Sheikhahmadi2024, tourchi2023temperature, Garakani2022}. While some clayey materials show only minor strength variation with temperature, others undergo significant softening, depending on the mineralogy, thermal history, and hydraulic boundary conditions. These complex, rate- and temperature-dependent behaviors underscore the importance of developing constitutive models that incorporate both thermally induced strength degradation and shear-rate sensitivity to reliably simulate long-term slope stability under thermal disturbances.

Several constitutive frameworks have been proposed to simulate these complex phenomena. Classical elasto-plastic models such as Mohr–Coulomb and Modified Cam-Clay, although widely used, are inadequate for capturing creep, thermal softening, or interface-localized failure \citep{Roscoe1968, Schofield1968}. Thermo-elasto-plastic extensions, including those developed by Hueckel and Baldi \citep{Hueckel1990} and Gens et al. \citep{Gens2009}, introduced temperature-dependent hardening and non-associated flow rules. More advanced formulations incorporate viscoplastic behavior using Perzyna-type overstress laws \citep{Perzyna1963, Borja1985}, which were later adapted for thermally driven creep in Boom clay by Cekerevac and Laloui \citep{Cekerevac2004}. Recent modeling efforts also highlight the need to treat slip surfaces as distinct interfaces, implemented using zero-thickness or double-node elements, to account for strain localization, anisotropic degradation, and discontinuous displacement fields \citep{Desrues1996, Dijkstra2010, Rafael2017}. Nevertheless, many models still treat residual strength as a fixed post-peak value, neglecting its evolution under temperature and rate-dependent deformation. Few formulations have successfully captured the progressive softening of residual strength with cumulative displacement, particularly in the presence of thermal loads \citep{Tang2016, Zhang2021}.

Temperature effects have been investigated in the context of soil shear strength. In drained testing environments, the overall shear strength of soil exhibits complex dependencies on temperature, often modulated by the accompanying volumetric changes. Drained triaxial compression tests on soft Bangkok clay showed that the drained shear strength increased as the soil temperature increased or after the specimen had been subjected to a temperature history \citep{abuel2007effect}. This increase in strength and stiffness is often attributed to thermal hardening \citep{burghignoli2000laboratory}, which results from the heating-induced contraction or permanent volume reduction typically observed in normally consolidated (NC) clays under drained conditions. This process generates an apparent preconsolidation pressure \citep{loche2025assessing}. In contrast, high temperatures can cause a decrease in strength in overconsolidated (OC) samples while inducing an increase in strength in NC samples \citep{scaringi2022a}. Studies on NC clay-concrete interfaces further showed that heating led to an improvement in shearing resistance and an increase in peak shear strength. In contrast, OC clays might exhibit thermal softening, where the induced reduction in shear strength becomes more pronounced at higher overconsolidation ratios \citep{yazdani2019influence}. 

\begin{figure}[H]
    \centering
    \includegraphics[width=0.9\linewidth]{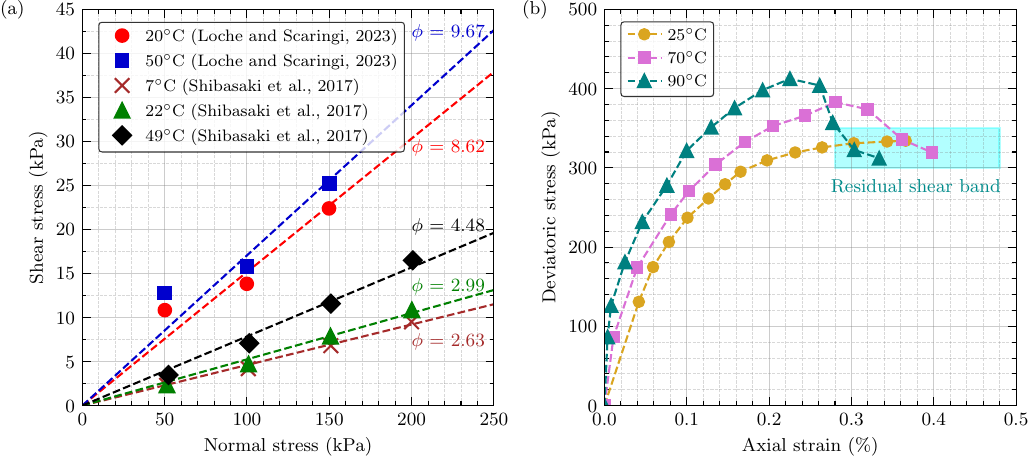}
    \caption{(a) Evolution of shear strength envelopes with temperature \citep{loche2023temperature, Shibasaki2017}, (b) Drained triaxial compression test results of normally consolidated clay specimens at different temperature levels \citep{abuel2007effect}.}
    \label{fig:stress-temp}
\end{figure}

The influence of thermal loading on shear strength can also be interpreted in terms of soil friction angle and cohesion. For normally consolidated clay--concrete interfaces, thermal loading was found to cause a significant increase in the interface friction angle \citep{yazdani2019influence}. Conversely, studies focusing on sand and kaolin clay over a temperature range of 5~$^{\circ}$C--40~$^{\circ}$C often reported that the effect of temperature on the friction angle was negligible \citep{yavari2016effect}. Figure \ref{fig:stress-temp}(a) indicates the direct relationship between temperature and shear strength envelopes in two different studies. The experimental results on high-plasticity soils show a significant dependence of soil shear strength on temperature. The findings from drained compression triaxial shear tests on specimens with a preconsolidation pressure of 300 kPa, conducted by \cite{abuel2007effect} on normally consolidated Bangkok clay and subjected to various temperatures (25, 70, and 90°C), are illustrated in Figure \ref{fig:stress-temp}(b). It is evident that specimens tested at higher temperatures achieved greater peak deviatoric stresses. Additionally, the normally consolidated specimens tested at room temperature (25°C) displayed strain hardening behavior, whereas those tested at elevated temperatures showed strain softening behavior. Figure \ref{fig:stress-temp}(b) shows that at large strains, the residual deviatoric stress of the soil remains unaffected by temperature. 

The influence of temperature on the effective cohesion ($c'$) or residual internal cohesion ($c_{res}$) is also context-dependent, though generally minimal in residual strength analyses \citep{shibasaki2016experimental, yavari2016effect, yazdani2019influence}. However, when considering soil/structure interfaces or initial peak strength, some temperature effects on adhesion (cohesion intercept) have been noted. For normally consolidated clay--concrete interfaces, thermal loading caused the interface adhesion to decrease slightly. Specifically, an interface adhesion decrease of 17\%--20\% was observed when the temperature increased from 24~$^{\circ}$C to 34~$^{\circ}$C \citep{yazdani2019influence}. Conversely, a separate study investigating 5~$^{\circ}$C to 40~$^{\circ}$C on clay and sand noted that the effect of temperature on cohesion was negligible \citep{yavari2016effect}.

Temperature variations, often overlooked in routine geotechnical practice, can significantly influence the stability of slopes, particularly those with pre-existing clay-rich slip surfaces. The previous studies confirm that soils experience temperature changes from various sources, including seasonal fluctuations, geothermal systems, and anthropogenic activities \citep{tanaka1997stress, burghignoli2000laboratory, loria2021thermally}. While daily temperature fluctuations may only penetrate a few decimeters, seasonal oscillations can affect the ground to depths of several meters, which is relevant for shallow landslides \citep{loche2022surface, jabbarzadeh2024thermo, tourchi2024thermo}. Field observations from landslide sites in temperate and cold regions have suggested a link between ground temperature changes and landslide activity, independent of rainfall or snowmelt triggers. For example, the Ru delle Roe landslide was triggered during an extremely cold period, which was hypothesized to have caused freezing of springs, leading to a buildup of pore water pressure \citep{govi1993example}. Similarly, the Touge landslide, a shallow slow-moving failure, consistently reactivates in the early cold season when ground temperatures decrease, a behavior that cannot be explained solely by pore pressure fluctuations \citep{shibasaki2016experimental}. These field observations underscore the need for models that can account for thermal effects on soil mechanical properties.  

To address these gaps, the present work introduces a non-isothermal viscoplastic constitutive model for interface elements, specifically designed to simulate the coupled thermal, rate-dependent, and irreversible behavior of clayey slip surfaces. The proposed model incorporates (i) temperature-dependent degradation of peak and residual strength, (ii) viscoplastic flow governed by a hyperbolic rate law, and (iii) stiffness reduction in both normal and shear directions as a function of temperature. The model is validated against ring-shear test data on bentonite and smectite-rich interfaces subjected to controlled thermal and mechanical loading paths. The proposed framework enables consistent modeling of thermal softening, time-dependent shear strain accumulation, and rate-dependent deformation relevant for long-term slope stability analysis under thermo-hydro-mechanical loading. The model is subsequently employed to analyze the Congress Street cut benchmark with the aim of evaluating how thermal cycling influences slope performance. The study highlights that thermally induced degradation of interface behavior can significantly diminish the effective stability of clayey slopes and, therefore, must be accounted for in stability evaluations subjected to temperature variations.

\section{Theoretical formulation}
\subsection {Non-isothermal elasto-viscoplastic model} 

Landslide slip surfaces, fault zones, and other localized failure mechanisms in clay-rich soils are often governed by highly nonlinear, path-dependent processes involving thermal, mechanical, and time-dependent effects. Experimental evidence has shown that temperature elevation accelerates strength degradation and stiffness loss, particularly in clays and weak rocks, leading to rapid shear localization. To capture this behavior, we developed a non-isothermal elasto-viscoplastic constitutive model for zero-thickness interface elements. The model was initially developed by \cite{alonso2013joints} and improved further in this study. Indeed, the enhanced model accounts for both temperature-dependent stiffness degradation and irreversible displacement accumulation, with the normal and tangential stiffnesses evolving as a function of thermal loading and shear rate. The framework employs a multiplicative decomposition of the relative displacement field into elastic and viscoplastic components and integrates thermally regulated softening laws to realistically simulate the progressive failure and healing behavior of shear bands. 

\subsubsection {Thermoelastic interface behavior}

The localized deformation observed in shear bands, slip surfaces, and fault zones is often modeled using zero-thickness interface elements, where relative displacement occurs due to discontinuous motion between adjacent solid phases. In this work, such elements are used to capture the onset and evolution of failure within materials like clays and weak rocks, particularly under thermo-mechanical loading conditions. The interface is formulated in a way that allows it to accommodate both elastic and irreversible viscoplastic displacements, with stiffnesses and strength evolving as a function of temperature and accumulated slip.

The displacement jump at the midpoint of the interface is interpolated from the nodal displacements of the surrounding finite elements. This interpolation accounts for the transformation between global and local coordinate systems. The temperature-dependent relative displacement vector $\boldsymbol{\delta}_{T,\text{mp}}$ at the midpoint of the interface is expressed as:

\begin{equation}
\boldsymbol{\delta}_{T,{mp}} = 
\begin{bmatrix}
\delta_{n,T} \\
\delta_{s,T}
\end{bmatrix}_{mp} = 
\boldsymbol{r} \boldsymbol{N_{mp}^{\delta}}
\begin{bmatrix}
-\boldsymbol{I}_4 & \boldsymbol{I}_4
\end{bmatrix}
\boldsymbol{u_j}
\end{equation}

\begin{figure}[H]
    \centering
    \includegraphics[width=1.0\linewidth]{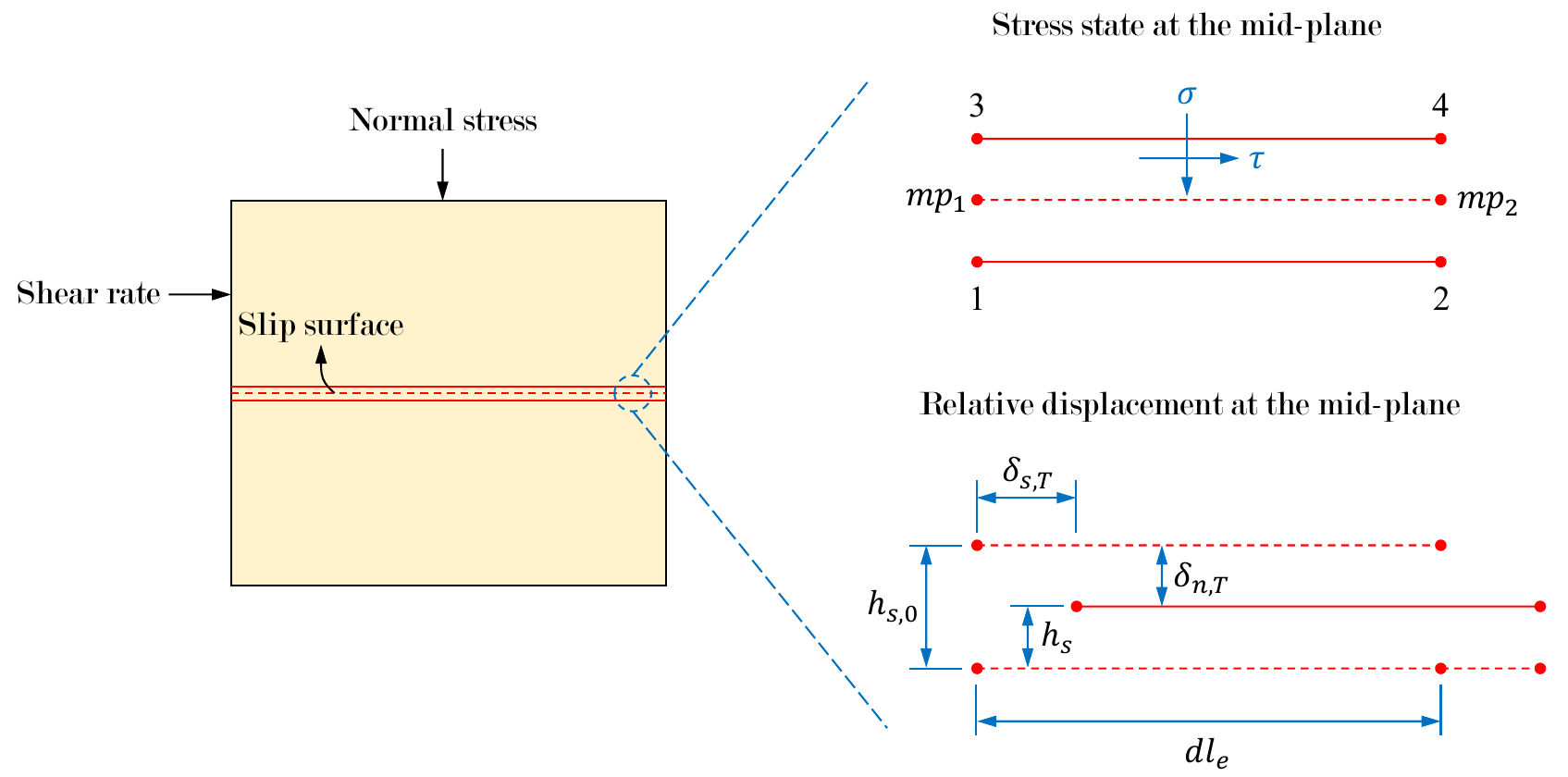}
    \caption{Interface element with double nodes.}
    \label{fig:interface_diagram}
\end{figure}

Here, $\delta_{n,T}$ and $\delta_{s,T}$ represent the temperature-dependent components of the relative displacement in the normal and tangential directions, respectively. The rotation matrix \(\boldsymbol{r}\) maps the local displacements to the global coordinate system, while the shape function matrix \(\boldsymbol{N_{mp}^{\delta}}\) is evaluated at the midpoint of the interface. The vector \(\boldsymbol{u_j}\) contains the nodal displacements associated with the interface element, and \(\begin{bmatrix} -\boldsymbol{I}_4 & \boldsymbol{I}_4 \end{bmatrix}\) is an identity matrix of 4th order. Figure \ref{fig:interface_diagram} is a conceptual representation of a zero-thickness interface element embedded along a slip surface in a shear test. The zoomed schematic illustrates the current and reference configurations used to define normal and tangential displacements \(\left(\delta_{n, T}, \delta_{s, T}\right)\), shear band thickness \(h_s\), and element length \(d l_e\).

The stress response of the interface is governed by a linear elastic law that links the temperature-dependent relative displacement to the effective normal and tangential stresses. This relationship is expressed as:

\begin{equation}
\boldsymbol{\sigma}'_{mp} = 
\begin{bmatrix}
\sigma' \\
\tau
\end{bmatrix}_{mp} = 
\boldsymbol{D}_T \, \boldsymbol{\delta}_{T,mp}
\end{equation}
where $\sigma'$ is the effective normal stress and $\tau$ is the tangential shear stress acting on the interface. The matrix $\boldsymbol{D}_T$ is the elastic stiffness matrix, which is temperature-dependent and defined by:

\begin{equation}
\boldsymbol{D}_T =
\begin{bmatrix}
K_{n,T} & 0 \\
0 & K_{s,T}
\end{bmatrix}
\end{equation}

In this formulation, $K_{n,T}$ and $K_{s,T}$ are the normal and tangential stiffnesses at temperature $T$, respectively. These values may evolve due to thermal effects and accumulated deformation. The effective stress is calculated by subtracting the maximum of the pore pressures acting within the interface from the total normal stress:

\begin{equation}
\boldsymbol {\sigma}'_{mp} = \boldsymbol {\sigma}_{mp} - \max(\boldsymbol{p}_{g,mp}, \boldsymbol {p}_{l,mp})
\end{equation}

Here, \(\boldsymbol {\sigma}_{mp}\) is the total normal stress, and \(\boldsymbol {p}_{g,mp}\), \(\boldsymbol {p}_{l,mp}\) represent the gas and liquid pore pressures, respectively. This formulation ensures that the interface can correctly simulate partial saturation and multiphase fluid effects in a coupled thermo-hydro-mechanical context.

The total relative displacement at the interface, $\boldsymbol{\delta}_T$, is decomposed into an elastic part $\boldsymbol{\delta}^e_T$ and a viscoplastic part $\boldsymbol{\delta}^{vp}_T$, which accounts for the irreversible slip along the interface. This decomposition is written as:

\begin{equation}
\boldsymbol{\delta}_T = \boldsymbol{\delta}^e_T + \boldsymbol{\delta}^{vp}_T
\end{equation}

The elastic part is recoverable and is directly related to the current stress state through the inverse of the stiffness matrix. The elastic displacement in each direction is given by:

\begin{equation}
\begin{bmatrix}
\delta^e_{n,T} \\
\delta^e_{s,T}
\end{bmatrix}
=
\begin{bmatrix}
1 / K_{n,T} & 0 \\
0 & 1 / K_{s,T}
\end{bmatrix}
\begin{bmatrix}
\sigma' \\
\tau
\end{bmatrix}
\end{equation}

This relation implies that stiffer interfaces will exhibit smaller elastic displacements under the same stress level, while thermally softened interfaces will allow larger elastic openings or slips.

To account for degradation and recovery of mechanical properties due to temperature and slip evolution, the stiffness parameters are modeled as functions of both the interface thickness and the current temperature. At the reference temperature $T_0$, the normal stiffness is defined by an inverse relationship with the effective thickness of the slip surface:

\begin{equation}
K_{n,T_0} = \frac{m}{h_s - h_{s,\min}}
\end{equation}

Here, $h_s$ is the instantaneous thickness of the shear band, $h_{s,\min}$ is a minimum admissible thickness used to prevent singularity, and $m$ is a scaling factor determined through calibration. This expression ensures that the interface stiffness becomes large as the shear band narrows, which is consistent with experimental observations in compacted clay specimens.

The actual stiffness at temperature $T$ is then modified to account for thermal softening or hardening depending on the direction and rate of viscoplastic slip. The temperature-dependent normal stiffness is written as:

\begin{equation}
K_{n,T} =
\begin{cases}
K_{n,T_0} \left[1 + \mu_{f_K} \ln \left( \dfrac{T}{T_0} \right) \right], & \dot{\delta}^{vp}_{n,T} < \left( \dot{\delta}^{vp}_{n,T} \right)_{\text{tr}} \\
K_{n,T_0}, & \dot{\delta}^{vp}_{n,T} = \left( \dot{\delta}^{vp}_{n,T} \right)_{\text{tr}} \\
K_{n,T_0} \left[1 - \mu_{f_K} \ln \left( \dfrac{T}{T_0} \right) \right], & \dot{\delta}^{vp}_{n,T} > \left( \dot{\delta}^{vp}_{n,T} \right)_{\text{tr}}
\end{cases}
\end{equation}

In this formulation, $\mu_{f_K}$ is a dimensionless parameter governing the rate of stiffness change with temperature, and $\left( \dot{\delta}^{vp}_{n,T} \right)_{\text{tr}}$ is a threshold rate below or above which softening or hardening behavior is activated. The model allows for both strengthening (under slow displacement rates) and softening (under rapid displacement) due to thermally induced changes in material bonding and microstructure.

The tangential stiffness $K_{s,T}$ is similarly defined:

\begin{equation}
K_{s,T} =
\begin{cases}
K_{s,T_0} \left[1 + \mu_{f_s} \ln \left( \dfrac{T}{T_0} \right) \right], & \dot{\delta}^{vp}_{s,T} < \left( \dot{\delta}^{vp}_{s,T} \right)_{\text{tr}} \\
K_{s,T_0}, & \dot{\delta}^{vp}_{s,T} = \left( \dot{\delta}^{vp}_{s,T} \right)_{\text{tr}} \\
K_{s,T_0} \left[1 - \mu_{f_s} \ln \left( \dfrac{T}{T_0} \right) \right], & \dot{\delta}^{vp}_{s,T} > \left( \dot{\delta}^{vp}_{s,T} \right)_{\text{tr}}
\end{cases}
\end{equation}

where $\mu_{f_s}$ is the tangential stiffness degradation parameter, and the remaining terms follow the same logic as the normal direction. These formulations ensure that the interface element can simulate realistic thermal degradation of stiffness and capture the transition from an intact bonded interface to a fully mobilized shear band with reduced resistance.

\subsubsection{Thermal yield surface and strength degradation}

The transition from purely elastic to inelastic response in the interface is governed by a yield surface that determines when viscoplastic slip is activated. For modeling of slip surfaces and shear zones in materials subjected to temperature variations, it is necessary to incorporate both temperature effects and progressive strength degradation into the yield criterion. The proposed formulation is therefore non-isothermal and includes temperature-dependent cohesion and friction angle, which evolve as a function of accumulated viscoplastic slip. The yield function is designed to be smooth and regularized, enabling robust numerical integration and avoiding discontinuities near yield onset.

The yield function $f_T$ is defined in a hyperbolic form, which allows a continuous transition from elastic to viscoplastic states and provides better numerical stability than piecewise or Mohr-Coulomb-like functions. It is expressed as:

\begin{equation}
f_T(\sigma', \tau) = \left( \frac{\tau}{q_{u,T}} \right)^2 + \alpha \left( \frac{\sigma'}{q_{u,T}} \right)^2 - 1
\end{equation}

In this equation, $\sigma'$ is the effective normal stress and $\tau$ is the shear stress acting at the interface. The parameter $\alpha$ controls the curvature of the yield function in the stress space and can be interpreted as a regularization factor that accounts for coupling between shear and normal stresses. The parameter $q_{u,T}$ represents the current temperature-dependent shear strength of the interface. This form of the yield function ensures that yielding occurs when the normalized combination of shear and normal stresses reaches unity.

The shear strength $q_{u,T}$ evolves as a function of the temperature-dependent cohesion $c_T$ and friction angle $\phi_T$ through the classical Mohr-Coulomb form:

\begin{equation}
q_{u,T} = c_T + \sigma' \tan \phi_T
\end{equation}

This relation reflects the combined contribution of apparent cohesion and frictional resistance to the overall shear strength. However, both $c_T$ and $\phi_T$ are no longer constant parameters; they degrade with progressive viscoplastic displacement and are influenced by temperature. To simulate softening behavior due to shear displacement accumulation, the cohesion is assumed to decrease linearly with respect to the irreversible tangential displacement:

\begin{equation}
c_T = c_{0,T} \left( 1 - \frac{\delta^{vp}_{s,T}}{\delta_c} \right)
\end{equation}

Here, $c_{0,T}$ is the initial temperature-dependent cohesion, $\delta^{vp}_{s,T}$ is the accumulated viscoplastic shear displacement, and $\delta_c$ is a characteristic displacement threshold at which the cohesion fully degrades. This linear softening model captures the progressive breakdown of bonding within the shear band as displacement continues to accumulate, which is commonly observed in post-peak behavior of clay interfaces. In parallel, the friction angle $\phi_T$ evolves from its peak value toward a residual value, following an exponential degradation law:

\begin{equation}
\tan \phi_T = \tan \phi_{res,T} + \left( \tan \phi_{peak,T} - \tan \phi_{res,T} \right) \exp\left( -\beta_d \delta^{vp}_{s,T} \right)
\end{equation}

This equation ensures a smooth and continuous transition in the friction angle as a function of shear displacement, where $\beta_d$ is a degradation parameter that controls the rate of softening. The use of the exponential function provides numerical stability and avoids abrupt changes in strength, which are known to cause convergence issues in finite element simulations. The residual friction angle $\phi_{res,T}$ itself is temperature-dependent, capturing the experimental observation that shear strength decreases with increasing temperature due to the weakening of particle contacts and loss of suction. This dependency is introduced using a logarithmic function:

\begin{equation}
\tan \phi_{res,T} = \tan \phi_{res,T_0} \left[ 1 - \xi_\phi \ln \left( \frac{T}{T_0} \right) \right]
\label{eq:temperature_dependent_interface_resistance}
\end{equation}

In this expression, $\xi_\phi$ is a temperature sensitivity coefficient, and $T_0$ is the reference temperature. A positive value of $\xi_\phi$ indicates that higher temperatures lead to lower residual strength, which is consistent with observed behavior in many thermo-sensitive soils.

The plastic admissibility conditions for the onset and continuation of viscoplastic flow are defined as:

\begin{equation}
f_T \leq 0, \quad \dot{f}_T = 0 \ \text{if} \ f_T = 0, \quad \dot{\delta}^{vp}_{s,T} \geq 0
\end{equation}

These conditions ensure that the material remains elastic when the stress state is inside the yield surface (i.e., $f_T < 0$), and viscoplastic slip evolves only when the yield surface is reached (i.e., $f_T = 0$) and the stress state remains on the yield surface during subsequent loading. Additionally, the condition $\dot{\delta}^{vp}_{s,T} \geq 0$ imposes irreversibility in the tangential slip direction, which is physically consistent with shear-induced softening processes in clays and weak interfaces.

This formulation enables the model to simulate the progressive loss of strength and stiffness observed in experimental studies on heat-activated slip surfaces and shear zones. It also allows the integration of thermal recovery or partial healing effects by adjusting the evolution laws for $c_T$ and $\phi_T$, should such mechanisms be relevant to the simulation scenario.

\subsubsection{Flow rule and viscoplastic slip evolution}

Once the yield condition is met, the interface undergoes irreversible deformation governed by a viscoplastic flow rule. In this section, we introduce a non-associated, rate-dependent flow formulation that is consistent with experimental observations of clay interfaces, where slip occurs gradually and under thermally influenced conditions. The formulation allows for both shear slip and normal dilation or contraction, depending on the local stress state and temperature.

The direction of viscoplastic flow is defined using a plastic potential function $G_T$, which governs the proportion of slip in the normal and tangential directions. The plastic flow direction is given by:

\begin{equation}
\dot{\boldsymbol{\delta}}^{vp}_T = \dot{\lambda}_T \frac{\partial G_T}{\partial \boldsymbol{\sigma}'}
\end{equation}
where $\dot{\lambda}_T$ is the temperature-dependent plastic multiplier, and $\partial G_T / \partial \boldsymbol{\sigma}'$ defines the flow direction in stress space. The use of a separate potential function $G_T \ne f_T$ enables a non-associated flow rule, which is necessary to reproduce the dilative or contractive behavior seen in interface shear tests.

To avoid numerical singularities and to capture the continuous evolution of deformation near yield, the plastic potential is defined in a smooth, hyperbolic form:

\begin{equation}
G_T(\sigma', \tau) = \sqrt{\tau^2 + \alpha^2 \sigma'^2} - \tau_g
\end{equation}

Here, $\alpha$ controls the coupling between shear and normal components, and $\tau_g$ is a regularization parameter that governs the shape and position of the flow potential. The square root form ensures differentiability and a smooth transition between sliding and sticking states. The parameter $\alpha$ may also be calibrated to reflect experimental data, such as dilation angles measured in shear tests at different confining pressures.

The evolution of viscoplastic slip is governed by a rate-dependent law derived from Perzyna-type viscoplasticity, which introduces a time scale into the constitutive response. The rate of plastic displacement is defined by:

\begin{equation}
\dot{\boldsymbol{\delta}}^{vp}_T = 
\begin{cases}
0, & f_T < 0 \\
\dot{\lambda}_T \dfrac{\partial G_T}{\partial \boldsymbol{\sigma}'}, & f_T = 0
\end{cases}
\end{equation}

This condition implies that no viscoplastic deformation occurs while the stress state lies strictly within the yield surface, and plastic flow is activated precisely at the yield threshold.

The rate of the plastic multiplier $\dot{\lambda}_T$ is expressed using a smooth overstress function:

\begin{equation}
\dot{\lambda}_T = \frac{1}{\eta} \left\langle \frac{f_T}{f_0} \right\rangle^n
\end{equation}

In this relation, $\eta$ is the viscosity coefficient controlling the rate sensitivity of the interface material, $f_0$ is a reference yield function value used to nondimensionalize the overstress, $n$ is the stress exponent that controls the nonlinearity of rate-dependence, and $\langle \cdot \rangle$ is the Macaulay bracket operator defined as $\langle x \rangle = \max(x, 0)$. This formulation ensures that the rate of plastic slip increases smoothly as the stress state exceeds the yield surface, and that it vanishes when $f_T < 0$, preserving the consistency of the elastic domain.

The evolution of the accumulated viscoplastic displacement $\delta^{vp}_{s,T}$ is updated at each time step by integrating the viscoplastic slip rate:

\begin{equation}
\delta^{vp}_{s,T}(t+\Delta t) = \delta^{vp}_{s,T}(t) + \dot{\delta}^{vp}_{s,T} \cdot \Delta t
\end{equation}

This cumulative displacement is used as an internal variable in the strength degradation functions for cohesion and friction angle, closing the loop between mechanical behavior, temperature effects, and material degradation. The proposed viscoplastic flow rule thus allows the model to capture realistic time-dependent interface behavior, including creep-like sliding under sustained load, rate effects during rapid shearing, and stiffness degradation due to temperature rise and accumulated slip.

The advective flux perpendicular to the interface is computed based on the transverse intrinsic permeability and the pressure gradient across opposing interface boundaries. In an analogous manner, advective transport along the interface relies on the intrinsic permeability in the longitudinal direction and is described using an extended version of Darcy's law. Consequently, it becomes crucial to specify intrinsic permeability values for both transverse and longitudinal directions of the interface. Furthermore, when dealing with unsaturated interfaces, defining the corresponding water retention curve is necessary.

The transversal flow is calculated as:

\begin{equation}
    q_l^t = \frac {k_l^t k_{rel}^t}{\mu_l} \breve{p}_{m p}
\end{equation}

\noindent where $k_l^t$ represents the intrinsic permeability of the liquid phase along the transversal direction, $k_{rel}^t$ corresponds to the relative permeability in the same direction, $\mu_l$ is the dynamic viscosity of the liquid, and $\breve{p}_{m p}$ is the pressure drop between the two surfaces of the joint element.

The generalized Darcy’s law for the longitudinal flow is also defined as:

\begin{equation}
    q_l^l = -\frac {k_l^l k_{rel}^l}{\mu_l} \left( \frac{\partial {p}_{mp}}{\partial l} - \rho \mathbf g \right)
\end{equation}

\noindent where $k_l^l$ is the longitudinal intrinsic permeability for the liquid, $k_{rel}^l$ is the longitudinal relative permeability for the liquid, ${p}_{mp}$ is the liquid pressure in the mid-plane, $\rho$ is the liquid density, and \(\mathbf g\) is the gravity vector. The non-advective flux, vapor diffusivity, is calculated using Fick's law:

\begin{equation}
    i_g^w = -\tau^* \rho_g S_g D_g^w \boldsymbol{I} \nabla \omega_g^w
\end{equation}

\noindent where $\tau^*$ denotes the tortuosity, $D_g^w$ is the molecular diffusion coefficient for water vapor in gas phase, which varies with temperature and gas pressure, \(\boldsymbol{I}\) is the identity matrix, $\nabla$ represents the gradient operator, and $\omega_g^w$ is the mass of water in gas phase.

The analysis of longitudinal fluid flow can be conducted under the assumption of laminar flow between two smooth and parallel plates separated by a defined hydraulic opening ($e$). Based on this assumption, the longitudinal hydraulic conductivity of the joint is determined using the cubic law:

\begin{equation}
    K_l = \frac{\rho \mathbf{g}}{\mu_l} \frac{e^3}{12}
\end{equation}

The intrinsic permeability can therefore be expressed as follows:

\begin{equation}
k_l^l = \frac{e^2}{12}
\end{equation}

The hydraulic aperture ($e$) at the interface can be connected to its geometric aperture ($h_s$) and joint roughness parameter ($r_i$), based on the relationship introduced by \cite{barton1986deformation}. By employing Barton’s formulation, the expression for the intrinsic permeability along the longitudinal direction becomes:

\begin{equation}
    k_l^l = \frac{1}{12} \left( \frac{h_s^2}{(r_i)^{2.5}} \right)^2
\end{equation}

The transverse intrinsic permeability $k_l^t$ is assumed to be equal to that of the surrounding porous medium.

The saturation level of the interface is determined using the standard retention curve suggested by 
\cite{vangenuchten1980}:

\begin{equation}
    S_l = \left[ 1 + \left( \frac{s}{P} \right)^{\frac{1}{1 - \lambda^*}} \right]^{-\lambda^*}
\end{equation}

\noindent where $s = P_g - P_l$ is the interface suction, $P_g$ and $P_l$ represent the gas and liquid pressures, respectively. The term $\lambda^*$ refers to a model-specific parameter, and $P$ denotes the air entry pressure required to initiate desaturation at the interface. The air entry pressure of the interface is influenced by the hydraulic aperture, as proposed by \citet{olivella2008gas}. This relationship can be interpreted through the Laplace equation, where the air entry pressure depends on the surface tension, contact angle, and the characteristic size of the aperture, and can be further linked to the permeability ratio between the current value ($k_l$) and reference value ($k_{l0}$):

\begin{equation}
    P = P_0 \sqrt{\frac{k_{l0}}{k_l}}
\end{equation}

Moreover, when temperature-dependent phenomena are accounted for, the value of $P$ is adjusted by the surface tension, $\sigma$:

\begin{equation}
    P = P_0 \sqrt{\frac{k_{l0}}{k_l}} \frac{\sigma}{\sigma_0}
\end{equation}

The relative permeability is calculated by

\begin{equation}
    k_{\text{rel}}^l = A S_l^n
\end{equation}

\noindent  where $n=3$ and $A=1$ can be considered for all cases.

Heat transfer is considered by the conduction mechanism using Fourier's law:

\begin{equation}
    i_C = -\lambda \nabla T
\end{equation}

In this context, $\lambda$ denotes the effective thermal conductivity of the interface element, and $\nabla T$ is the temperature gradient. To account for the influence of moisture content, $\lambda$ can be modeled as a function of the liquid saturation degree $S_l$:

\begin{equation}
\lambda = \lambda_{\text{sat}} \sqrt{S_l} + \lambda_{\text{dry}} (1 - \sqrt{S_l})
\end{equation}

\noindent where, $\lambda_{\text{sat}}$ and $\lambda_{\text{dry}}$ represent the thermal conductivities under saturated and dry conditions, respectively.

\subsection{Balance equations}
\subsubsection{Water mass balance equation}
The behavior of two-phase flow within the interface elements is characterized by deriving the governing equations for mass and energy conservation of water and air at the interface. Water may exist in both the liquid and vapor phases. The overall mass balance of water for a differential segment of the interface element is given as:

\begin{equation}
    \frac{\partial(\theta_l^w S_l + \theta_g^w S_g)}{\partial t} \, h_s \, dL_e 
    + (\theta_l^w S_l + \theta_g^w S_g) \frac{dh_s}{dt} \, dL_e 
    + [j_l^{\prime w}]_{mp} 
    + [j_g^{\prime w}]_{mp} 
    = f^w
\end{equation}

\noindent where $\theta_l^w$ and $\theta_g^w$ represent the water content in the liquid and vapor phases, respectively. The parameter $dL_e$ is the discrete length of the interface element. The saturation degrees in the liquid and gas phases are denoted by $S_l$ and $S_g$, respectively. The terms $\left[j_l^{\prime w}\right]_{mp}$ and $\left[j_g^{\prime w}\right]_{mp}$ correspond to the fluxes of liquid and vapor water through the interface, while $f^w$ denotes an external source term contributing to water mass.

\begin{equation}
    [j_l^{\prime w}]_{mp} = 
    [\theta_l^w q_l^t \, dL_e + i_{l}^{w,t} \, dL_e]_{0}^{h_s} 
    + [\theta_l^w q_l^l \, h_s + i_{l}^{w,l} \, h_s]_{0}^{dL_e}
\end{equation}

\begin{equation}
    [j_g^{\prime w}]_{mp} = 
    [\theta_g^w q_g^t \, dL_e + i_{g}^{w,t} \, dL_e]_{0}^{h_s} 
    + [\theta_g^w q_g^l \, h_s + i_{g}^{w,l} \, h_s]_{0}^{dL_e}
\end{equation}

\noindent where $q_l^t$, $q_g^t$, $q_l^l$, and $q_g^l$ denote the advective fluxes (for liquid and gas phases) in the transverse and longitudinal directions at the boundaries of the interface element. Similarly, $i_{l}^{w,t}$, $i_{g}^{w,t}$, $i_{l}^{w,l}$, and $i_{g}^{w,l}$ represent the transverse and longitudinal non-advective flux components, respectively. 

\subsubsection{Air mass balance equation}
The air mass balance formulation accounts for both free air and air dissolved in the liquid phase. Its mathematical form is as follows:

\begin{equation}
    \frac{\partial(\theta_l^a S_l + \theta_g^a S_g)}{\partial t} \, h_s \, dL_e 
    + (\theta_l^a S_l + \theta_g^a S_g) \frac{dh_s}{dt} \, dL_e 
    + [j_l^{\prime a}]_{mp} 
    + [j_g^{\prime a}]_{mp} 
    = f^a
\end{equation}

\noindent where $\theta_l^a$ represents the air content dissolved in the liquid phase, and $\theta_g^a$ refers to the amount of air in the gas (dry) phase. The term $\left[j_l^{\prime a}\right]_{mp}$ corresponds to the dissolved air flux at the interface, while $\left[j_g^{\prime a}\right]_{mp}$ denotes the gas-phase air flux. The parameter $f^a$ accounts for any external input of air to the system.

\subsubsection{Internal energy balance equation}
The internal energy balance for the interface element is expressed by:

\begin{equation}
    \frac{\partial(E_l \rho_l S_l + E_g \rho_g S_g)}{\partial t} \, h_s \, dL_e 
    + (E_l \rho_l S_l + E_g \rho_g S_g) \frac{dh_s}{dt} \, dL_e 
    + [i_c]_{mp} 
    + [j_{El}]_{mp} 
    + [j_{Eg}]_{mp} 
    = f^E
\end{equation}

The energies of the liquid and gas phases are computed by:

\begin{equation}
    E_l \rho_l = (E_l^w \omega_l^w + E_l^a \omega_l^a) \rho_l = E_l^w \theta_l^w + E_l^a \theta_l^a
\end{equation}

\begin{equation}
    E_g \rho_g = (E_g^w \omega_g^w + E_g^a \omega_g^a) \rho_g = E_g^w \theta_g^w + E_g^a \theta_g^a
\end{equation}

\noindent where $E_l^w$ and $E_l^a$ refer to the internal energies (per unit mass) of water and air within the liquid. The $\omega_l^w$ and $\omega_l^a$ indicate the corresponding masses of water and air in the liquid phase. Similarly, $E_g^w$ and $E_g^a$ denote the internal energies of water and air in the gas phase, while $\omega_g^w$ and $\omega_g^a$ represent their respective masses. The densities of the liquid and gas phases are denoted by $\rho_l$ and $\rho_g$, respectively. The internal energies of the liquid and gas phases are denoted by $E_l$ and $E_g$. The term $[i_c]_{mp}$ corresponds to the conductive heat flux through the zero-thickness interface element. The quantities $[j_{E_l}]_{mp}$ and $[j_{E_g}]_{mp}$ represent the advective energy fluxes associated with the liquid and gas phases at the interface. The heat conduction at mid-plane of interface is defined as:

\begin{equation}
    [i_c]_{mp} = [i_c^t \, dL_e]_{0}^{h_s} + [i_c^l \, h_s]_{0}^{dL_e}
\end{equation}

\noindent where $i_c^t$ and $i_c^l$ represent the heat flux components in the transverse and longitudinal directions at the boundaries of the interface element, respectively.

The quantities $[j_{E_l}]_{mp}$ and $[j_{E_g}]_{mp}$ are determined considering the advective transport of energy, as follows:

\begin{equation}
    [j_{El}]_{mp} = [j_l^{\prime w}]_{mp} E_l^w + [j_l^{\prime a}]_{mp} E_l^a
\end{equation}

\begin{equation}
    [j_{Eg}]_{mp} = [j_g^{\prime w}]_{mp} E_g^w + [j_g^{\prime a}]_{mp} E_g^a
\end{equation}

The balance equations are solved in a coupled manner. The primary unknowns at each node consist of the normal and tangential relative displacements ($\delta_{n,T}$, $\delta_{s,T}$), gas and liquid pressures ($P_g$, $P_l$), and temperature $T$. Advective flow along the longitudinal direction is governed by Darcy’s law. Diffusive transport, or non-advective fluxes, is represented using Fick’s law. The interface’s hydraulic properties, including permeability and air entry pressure, are influenced by the aperture size. Heat conduction across the interface is evaluated according to Fourier’s law. The mechanical behavior of the interface is considered using a non-isothermal elasto-viscoplastic framework, described in Section 2.1. 

\section{Model validation}
\subsection{Ring-shear test}
\begin{figure}[H]
    \centering
    \includegraphics[width=0.55\linewidth]{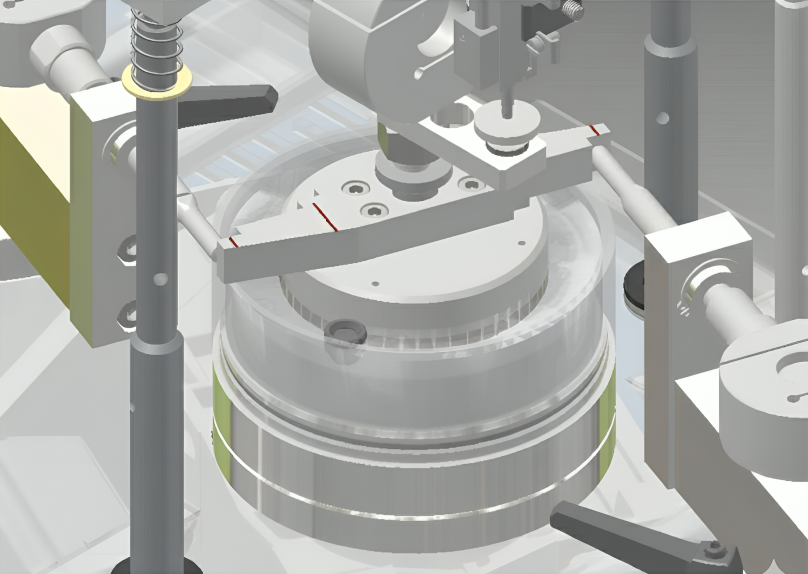}
    \caption{A three-dimensional schematic illustration of a ring-shear device (after \cite{loche2023temperature}).}
    \label{fig:schematic_ring}
\end{figure}

The ring-shear test stands as a critical experimental methodology in geotechnical and geological hazard analysis, particularly recognized as the most suitable technique for accurately measuring the residual strength of soils \citep{Shibasaki2017}. A typical 3D schematic illustration of a ring-shear device is presented in Figure \ref{fig:schematic_ring}. A key advantage of this test over traditional shear tests is its ability to allow for large displacements along a single direction while ensuring that the shear plane area remains constant during shearing, thereby permitting unlimited continuous shear deformation of the specimen \citep{li2013ring}. This capability is essential for simulating long sliding and shearing courses, such as those involved in landslides and debris flows. The test is fundamental for characterizing strain softening behavior and residual strength characteristics of soils under large displacements. The resulting residual strength is vital for evaluating the long-term stability and potential reactivation of landslides.

\begin{figure}[H]
    \centering
    \includegraphics[width=0.7\linewidth]{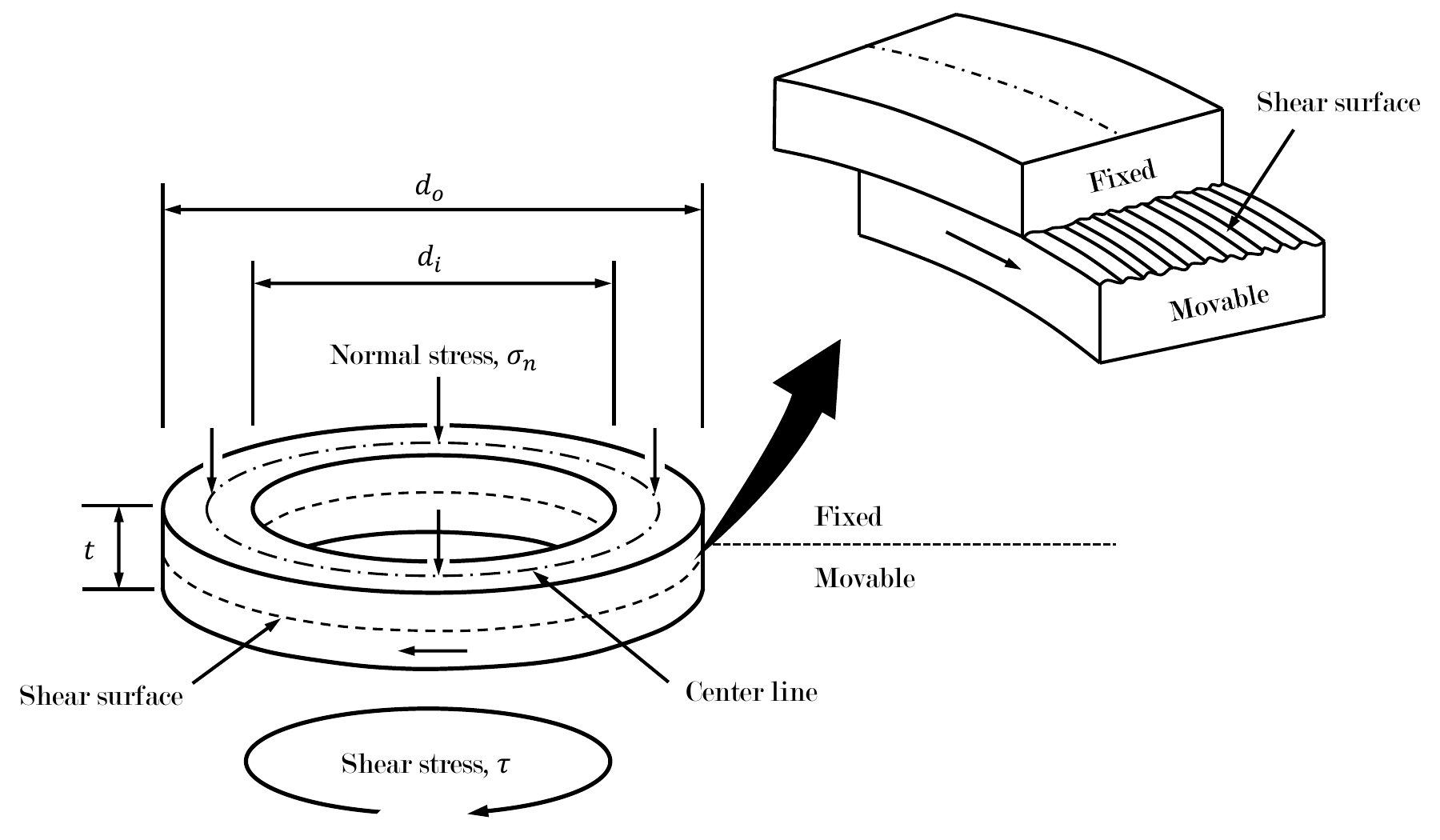}
    \caption{Schematic illustration of the test specimen during a ring-shear experiment.}
    \label{fig:test_specimen}
\end{figure}

The ring-shear test is conducted on an annular ring-shaped specimen which is laterally confined and subjected to a constant normal stress. Shear deformation is induced by the relative rotary motion of the specimen \citep{zhang2011ring}. This is indicated schematically in the Figure \ref{fig:test_specimen}. The shear stress is generally assumed to be uniformly applied on the shearing surface. \cite{bishop1971new} suggested that this assumption introduces an insignificant error in the calculation of normal stress ($\sigma_n$) and shear stress ($\tau$). Ring-shear tests can be conducted under different displacement- or rate-controlled conditions, with the shearing process managed by adjusting torque or shear rate. The test can be carried out under drained (e.g., simulating naturally draining slip zones where pipes are open during shearing) or undrained conditions (simulating scenarios where pore water cannot rapidly leave the soil mass compared to volume and water pressure changes). Following shearing, the strength gradually decreases to a residual value, a phenomenon often attributed to the reorientation of clay mineral particles parallel to the shearing direction \citep{ma2019investigation}.

The temperature-controlled ring-shear tests are essential for exploring the thermo-mechanical behavior of geomaterials, particularly clayey soils. Experimental results show that the shear strength characteristics of clays are temperature-dependent, exhibiting shear-weakening behavior with decreasing temperature under slow shearing rates (e.g., $\le 0.1\text{ mm/min}$). This finding is particularly significant for understanding mechanisms of shallow, slow-moving landslides that may be activated by seasonal decreases in ground temperature, a phenomenon observed in field studies \citep{shibasaki2016experimental}. Conversely, at higher shearing rates, this thermal effect tends to disappear or reverse, leading to strength gain with decreasing temperature, possibly due to increased viscous resistance from particle disorder \citep{scaringi2022a}.

In this study, the proposed model was validated against experimental measurements from temperature-controlled drained ring-shear tests conducted under three different thermal loading scenarios. In the first scenario, the soil sample initially reaches its residual shear strength and then undergoes heating. Afterward, the temperature of the soil sample decreases back to the initial room temperature, following a cooling path. This is referred to as the heating-cooling scenario. In the second scenario, the process is reversed: the soil undergoes cooling first, followed by heating along a cooling-heating path. The third scenario combines elements from the first two. After reaching the residual shear strength, the soil first follows the heating-cooling path and then shifts to the cooling-heating path. These experimental scenarios were intentionally selected from the literature to assess the model's capability under various thermal loading conditions.

\subsection{Numerical simulation of heating-cooling scenario}

\begin{figure}[H]
    \centering
    \includegraphics[width=0.55\linewidth]{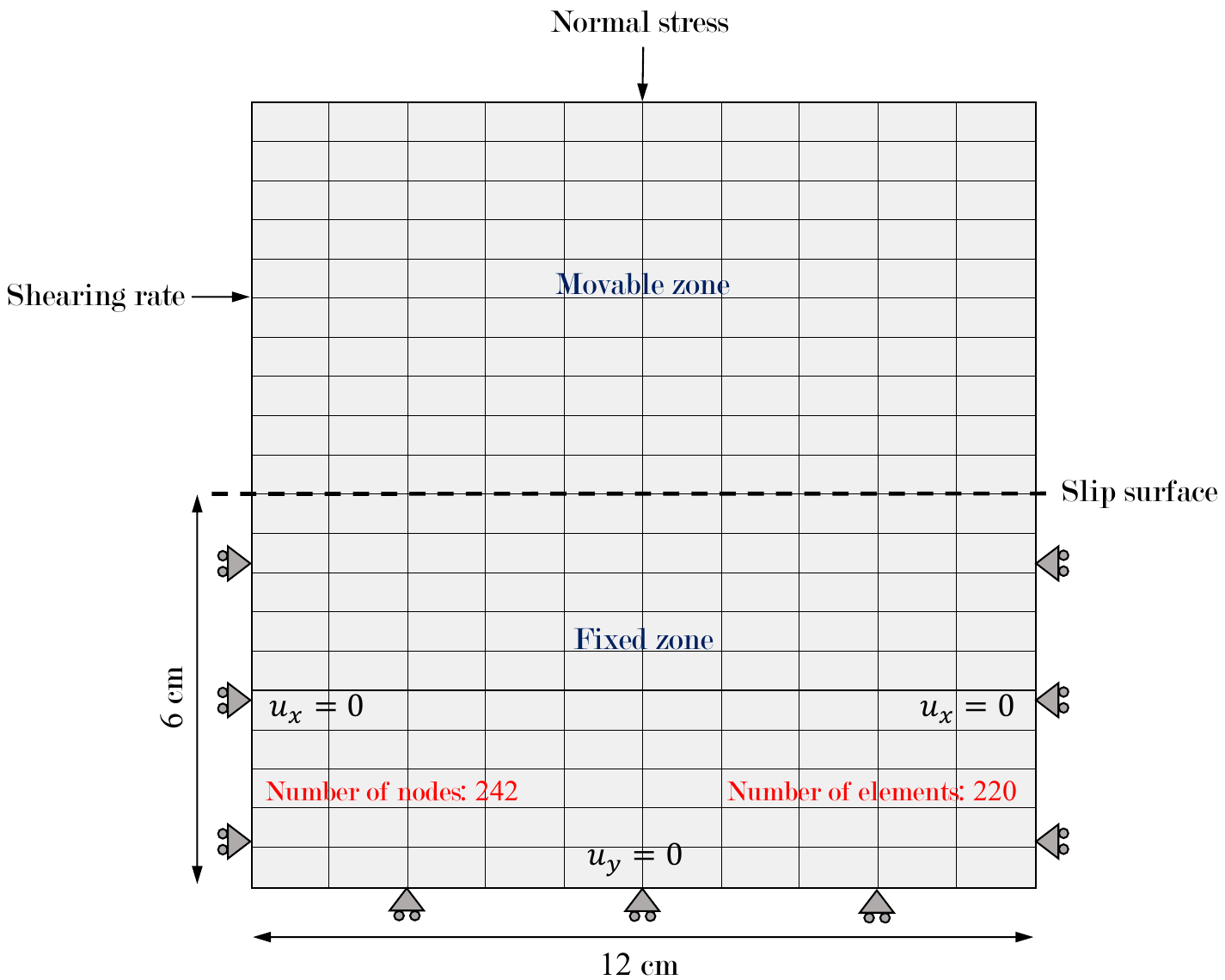}
    \caption{Model geometry, boundary conditions, and finite-element mesh used in numerical simulations.}
    \label{fig:Geometry}
\end{figure}

First, the model is validated using experimental ring-shear test results from \cite{loche2023temperature} conducted under the heating-cooling path. A commercially available bentonite was employed for temperature-controlled ring-shear tests. The material, known as the Czech B75 Bentonite, is a Ca-Mg-bentonite extracted from a deposit in Černý vrch, Czech Republic. Some of the soil properties are listed in Table \ref{tab:bentonite_properties}. Further characterization can be found in \citep{sun2020}. The model geometry, mechanical boundary conditions, and finite-element mesh are illustrated in Figure \ref{fig:Geometry}.

\begin{table}[h]
    \centering
    \caption{Basic properties of the tested bentonite.}
    \begin{tabular}{l|c|l|c}
        \hline
        \textbf{Parameter} & \textbf{Value} & \textbf{Parameter} & \textbf{Value} \\
        \hline
        Specific gravity, \(G_s\) & 2.87 & Activity, \(A\) & 2.7 \\
        Liquid limit, \(LL\) & 217 & Clay fraction, \(c_f\) & 61 \\
        Plastic limit, \(PL\) & 51 & Silt fraction, \(s_f\) & 33 \\
        Ca-Mg montmorillonite & 85 & USCS name & CH \\
        \hline
    \end{tabular}
    \label{tab:bentonite_properties}
\end{table}

The tests were performed in a conventional Bromhead-type ring-shear apparatus \citep{bromhead1979} equipped with a temperature-change device allowing water circulation in a closed circuit between an external temperature-controlled bath and the shear-box bath. The device accommodates a 5 mm-thick annular sample sandwiched between brass porous platens that are roughened to avoid interface shearing. Lateral friction was minimized by ensuring post-consolidation sample thicknesses larger than 4.25 mm. The available range of shear rates was exploited (\(v = 0.018 - 44.5 \, \text{mm/min}\)), which are associated with slow-to-rapid landslide movements. The samples were reconstituted following \cite{burland1990} and consolidated stepwise to \(\sigma_v' = 600 \, \text{kPa}\). They were then unloaded to \(\sigma_v' = 50 - 150 \, \text{kPa}\) before shearing. Under each \(\sigma_v'\) level, \(v\) was increased stepwise. A schematic illustration of the test specimen is shown in Figure \ref{fig:test_specimen}. Further details can be found in \cite{loche2023temperature}.

\begin{figure}[H]
    \centering
    \includegraphics[width=0.5\linewidth]{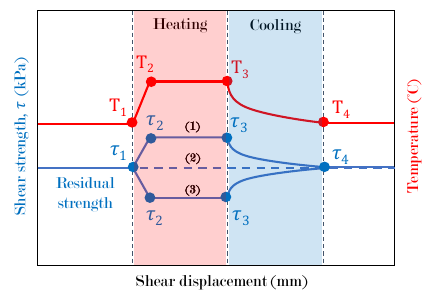}
    \caption{Schematic illustration of temperature-controlled ring-shear tests.}
    \label{fig:figure2}
\end{figure}

\begin{figure}[H]

  \centering
  \includegraphics[width=1.0\textwidth]{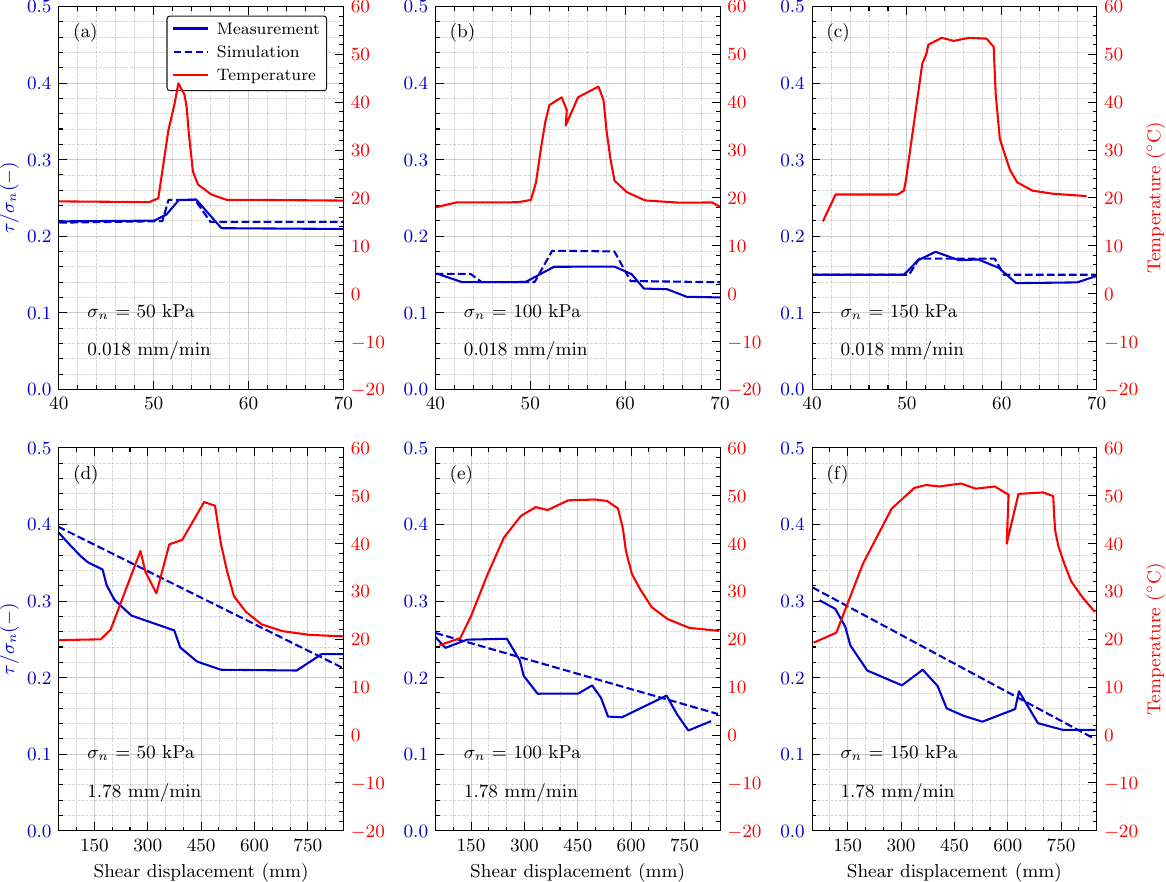} \\ \vspace{0.5cm}
  \caption{The variations in shear resistance under heating-cooling paths during ring-shear tests.}
  \label{fig:figure3}
\end{figure}

Following the path shown in Figure~\ref{fig:figure2}, heating-cooling tests were conducted. As illustrated in Figure~\ref{fig:figure2}, under such conditions, the shear strength may increase, decrease, or remain unchanged. This can be related to the rate-dependent behavior. In the experiments, after reaching the residual shear strength under the selected stress and displacement‐rate conditions at room temperature (\(20\,^\circ\mathrm{C}\)), the bath temperature was raised to \(55\,^\circ\mathrm{C}\) and held constant over a sufficient shearing distance before being returned to the initial temperature. Under slow shearing (\(0.018\,\mathrm{mm/min}\)), shear resistance increased with temperature (Figures~\ref{fig:figure3}a–c). At a moderate rate (\(1.78\,\mathrm{mm/min}\)), that trend was not clear, and some thermal weakening was detected (Figures~\ref{fig:figure3}d–f). At an intermediate rate (\(v = 0.5\,\mathrm{mm/min}\)), strength changes were minimal, suggesting this rate may mark an upper bound for thermal strengthening and may correspond to typical landslide speeds. Tests at \(v = 0.5\,\mathrm{mm/min}\) and at a high rate (\(44.5\,\mathrm{mm/min}\)) are not considered here, since maintaining drained conditions at high rates can be difficult.

The ring-shear tests with rates of \(0.018 \, \text{mm/min}\) and \(1.78 \, \text{mm/min}\) were selected to check the capabilities of the proposed model. The numerical simulation was carried out with the developed new non-isothermal constitutive model formulated above. The simulation is assumed under two-dimensional coupled THM conditions. The pore pressure dissipates during drained experiments, and the simulation aims to capture the soil thermo-mechanical response. The rate of displacements used in the test is applied to the movable part. The fixed part is constrained horizontally and vertically at the bottom. The net normal stresses used in the test are applied to the top of the movable part. The interface element is discretized into 20 elements. An initial constant temperature of \(21.8 \, ^\circ C\) has been assumed throughout the geometry. Heat power was applied as a thermal flux on the interface. The fixed and movable parts (Figure \ref{fig:test_specimen}) are represented as elastic materials, and the shear interface is modeled as a visco-plastic zero-thickness element. The parameters are detailed in Table \ref{tab:parameters}.

The predictions from the numerical analysis are plotted as a dashed line alongside the test measurements in Figure \ref{fig:figure3}. The results demonstrate that the material's strength behavior varies significantly with the shearing rate. At slow rates, an increase in strength was observed with temperature, while moderate rates caused weakening. Interestingly, negligible changes were seen at an intermediate rate, suggesting an upper limit for thermal strengthening. The numerical analysis has simulated the various phases of the experiment, as shown in Figure \ref{fig:figure2}. It is evident that the overall shear behavior is reasonably well captured by the model and the calculated results are in good agreement with the measured values. 

\begin{table}[H]
    \centering
    \caption{Model parameters for simulating the ring-shear tests by \cite{loche2023temperature}.}
    \begin{tabular}{l|l|l}
        \hline
        \textbf{Parameter} & \textbf{Value} & \textbf{Unit} \\
        \hline
        Mechanical properties & & \\
        \hline
        Initial normal stiffness parameter, \(m\) & \(60 \times 10^3\) & kPa \\
        Tangential stiffness, \(K_s^{T_0}\) & \(166 \times 10^3\) & kPa/m \\
        Initial friction angle, \(\phi_0\) & 35 & \( ^\circ \) \\
        Residual friction angle, \(\phi_{\text{res}}^{(T_0)}\) & 8.69 & \( ^\circ \) \\
        & 12.18 & \( ^\circ \) \\
        & 16 & \\
        Initial shear surface thickness, \(t_0\) & 0.1 & mm \\
        Minimum shear surface thickness, \(h_{s,\min}\) & 0.01 & mm \\
        Viscosity, \(\eta_v\) & \(1 \times 10^{-2}\) & s\(^{-1}\) \\
        Stress power, \(m_v\) & 2.0 & - \\
        Critical displacement for cohesion, \(u_c^*\) & 1 & mm \\
        Critical displacement for \(\tan \phi\) & 1 & mm \\
        Uniaxial compressive strength, \(q_u\) & 600 & kPa \\
        Model parameter, \(\beta_d\) & 100 & - \\
        \hline
        Non-isothermal parameters & & \\
        \hline
        Residual friction angle, \(\phi_{\text{res}}^T\) & 9.85 & \( ^\circ \) \\
        & 11.69 & \( ^\circ \) \\
        & 8.53 & \( ^\circ \) \\
        Model parameter, \(\xi_\phi\) & 0.5 & - \\
        \hline
    \end{tabular}
    \label{tab:parameters}
\end{table}

\subsection{Numerical simulation of cooling-heating and combined scenarios}

For further model validation experimental data from a comprehensive study by \cite{Shibasaki2017} on the temperature-dependent residual shear strength of landslide soils were utilized. They employed ring-shear tests to investigate the effects of temperature on 23 different soil samples, including natural landslide soils and commercial clays, with varying smectite contents. The experimental results selected for model validation encompass both cooling-heating paths and a combined cooling-heating and heating-cooling path, providing a complete dataset to test the model's predictive capabilities under different thermal loading scenarios. 
The validation of the model under a cooling-heating path was performed using the results for four specific smectite-bearing soils, identified as samples No. 1, 2, 3, and 5 in the original study. These soils were selected for their distinct properties and high smectite content, which makes their shear strength sensitive to temperature changes. For the combined scenario, data from Sample No. 6 was used. The physical properties for each soil sample are reported in Table \ref{tab:soil_properties}.

\begin{table}[H]
  \centering
  \caption{The physical properties of the soil samples}
  \label{tab:soil_properties}
  \begin{tabular}{l l l l l l l l}
    \toprule
    \multirow{2}{*}{No.} & \multicolumn{3}{c}{Index properties (\%)} & \multicolumn{3}{c}{Grain size (\%)} & \multirow{2}{*}{Smectite fraction (\%)} \\
    \cmidrule(lr){2-4} \cmidrule(lr){5-7}
     & LL & PL & PI & Clay & Silt & Sand & \\
    \midrule
    1 & 162.9 & 46.3 & 116.6 & 48 & 52 & 0  & 64 \\
    2 & 131.2 & 49.9 & 81.3  & 32 & 68 & 0  & 64 \\
    3 & 121.9 & 41.5 & 80.4  & 66 & 27 & 7  & 50 \\
    5 & 223.6 & 78.3 & 145.3 & 24 & 76 & 0  & 44 \\
    6 & 144.8 & 37.7 & 107.1 & 53 & 36 & 11 & 42 \\
    \bottomrule
  \end{tabular}
  \vspace{2mm}
\end{table}

The general procedure involved shearing the samples at a slow and constant velocity until a steady residual strength was achieved at room temperature. After this, a cooling event was initiated by either adding ice to the shear-box bath or circulating temperature-controlled water. The tests were performed under drained conditions in ring-shear apparatuses, with a small gap maintained between the confining rings to prevent friction and allow drainage. All selected experiments were performed under a normal stress of 200 kPa. For samples No. 1, 2, and 5, a constant velocity of 0.005 mm/min was maintained, while for samples No. 3 and 6, shear rates of 0.05 mm/min and 0.02 mm/min were applied, respectively. The model parameters used for simulations are reported in Table \ref{tab:soil_model_properties}.

\begin{table}[htbp]
  \centering
  \caption{Model parameters for simulating the ring-shear tests by \cite{Shibasaki2017}.}
  \label{tab:soil_model_properties}
  \renewcommand{\arraystretch}{1.2}
  \resizebox{\textwidth}{!}{%
  \begin{tabular}{llllll}
    \toprule
    \textbf{Parameter} & \textbf{No. 1} & \textbf{No. 2} & \textbf{No. 3} & \textbf{No. 5} & \textbf{No. 6} \\
    \midrule
    Initial stiffness $m$ (kPa)              & $30 \times 10^3$ & $25 \times 10^3$ & $35 \times 10^3$ & $60 \times 10^3$ & $40 \times 10^3$ \\
    Tangential stiffness $K_s^{T_0}$ (kPa/m)  & $80 \times 10^3$ & $70 \times 10^3$ & $90 \times 10^3$ & $166 \times 10^3$ & $100 \times 10^3$ \\
    Initial friction angle $\phi_0$ ($^\circ$) & 30 & 29 & 31 & 35 & 32 \\
    \multirow[t]{3}{*}{Residual friction angle $\phi_{res}^{(T)}$ ($^\circ$)} 
      & 14.0 & 13.0 & 15.0 & 12.2 & 16.0 \\ 
      & 10.5 & 9.8  & 11.5 & 8.7  & 12.5 \\ 
      & 13.8 & 12.7 & 14.6 & 11.9 & 15.5 \\
    Shear thickness $t_0$ (mm)               & 0.10 & 0.10 & 0.10 & 0.10 & 0.10 \\
    Minimum thickness $h_{s,\min}$ (mm)      & 0.01 & 0.01 & 0.01 & 0.01 & 0.01 \\
    Viscosity $\eta_v$ (s$^{-1}$)            & $2.0 \times 10^{-2}$ & $1.5 \times 10^{-2}$ & $2.0 \times 10^{-2}$ & $1.0 \times 10^{-2}$ & $2.0 \times 10^{-2}$ \\
    Stress power $m_v$ (-)                   & 2.0 & 2.0 & 2.0 & 2.0 & 2.0 \\
    Critical disp. for cohesion $u_c^*$ (mm) & 1.0 & 1.0 & 1.0 & 1.0 & 1.0 \\
    Critical disp. for $\tan\phi$ (mm)       & 1.0 & 1.0 & 1.0 & 1.0 & 1.0 \\
    Uniaxial comp. strength $q_u$ (kPa)      & 500 & 480 & 550 & 600 & 520 \\
    Model parameter $\beta_d$ (-)            & 100 & 100 & 100 & 100 & 100 \\
    Non-isothermal param. $\xi_\phi$ (-)     & 0.5 & 0.5 & 0.5 & 0.5 & 0.5 \\
    \bottomrule
  \end{tabular}
  }
\end{table}

\begin{figure}[H]
  \centering
  \includegraphics[width=0.8\textwidth]{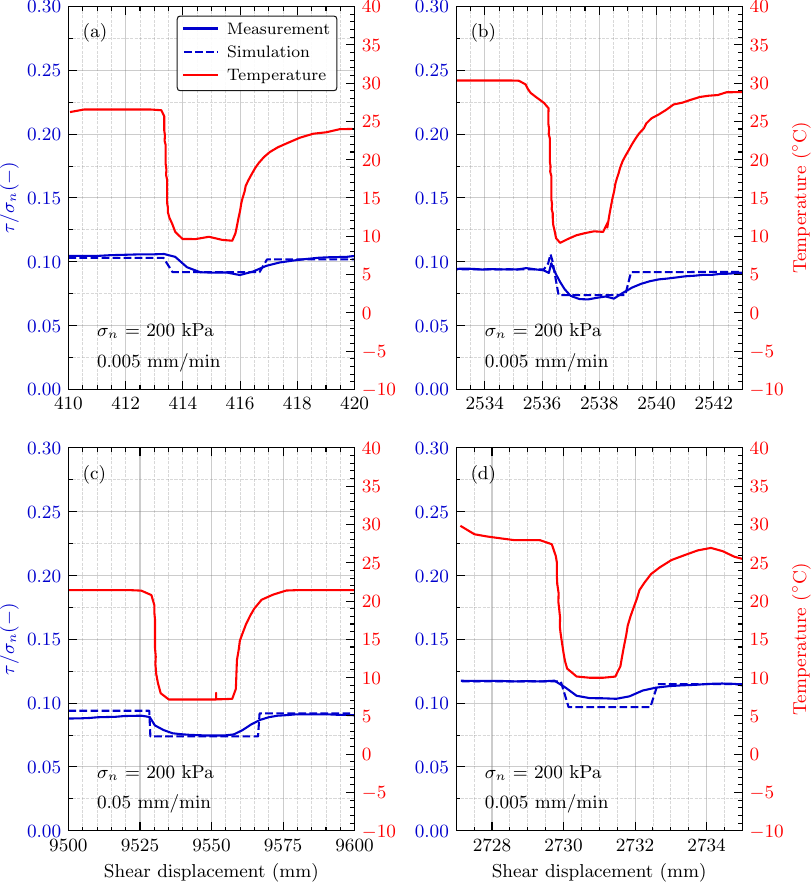}
  \caption{The variations in shear resistance under cooling-heating paths during ring-shear tests: sample (a) No. 1, (b) No. 2, (c) No.3, and (d) No. 5.}
  \label{fig:figure_5}
\end{figure}

\begin{figure}[H]
  \centering
  \includegraphics[width=0.5\textwidth]{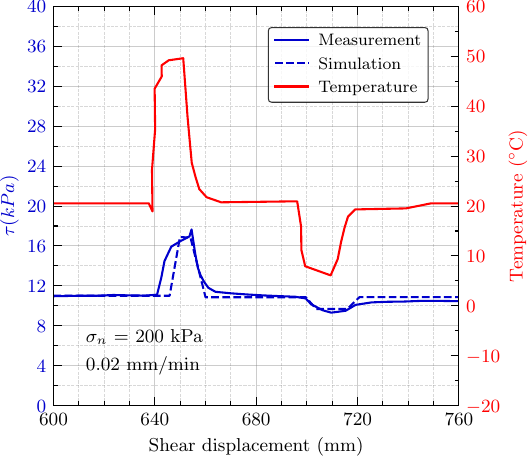}
  \caption{The variations in shear stress for sample No. 6 under combined thermal path during ring-shear tests.}
  \label{fig:figure_6}
\end{figure}

The comparison between the experimental results and model simulations along the cooling-heating path is illustrated in Figure \ref{fig:figure_5}. A clear reduction in shear resistance is observed in both the experiments and simulations during cooling. This reduction is attributed to shear weakening resulting from the decrease in soil temperature, which is modeled using the temperature-dependent soil stiffness equations (Eqs. 8 and 9) proposed in this study. The model is also capable of simulating the effects of combined thermal cycles on the shear resistance of the soil. This is supported by the comparison shown in Figure \ref{fig:figure_6}. The simulation results for a combined thermal path align well with the experimental findings.

\section{Case study}
\label{sec:case_study}

\subsection{Congress Street cut}
\label{subsec:congress_street_cut}

The Congress Street cut in Chicago is a classical and extensively documented case history widely used for validating numerical models in slope stability analysis \citep{Ireland1954, Oka1990, Ching2009, Ji2012, Li2016, Jiang2016, Li2019, Wang2020}. The failure occurred in 1952 during the excavation of an open cut approximately 60~m long, located on the south side of the project. The subsurface profile consists of a relatively thin layer of sand and miscellaneous fill overlying three distinct layers of glacial clay. The upper clay layer is stiff and strongly affected by desiccation, leading to fissures, joints, and cracks. In contrast, the two underlying clay layers are medium stiff and more homogeneous. A deeper, stiffer clay stratum limits the downward extension of the failure surface.

The observed failure mechanism was predominantly rotational, with a circular slip surface developing within the upper clay layers and extending toward the slope toe. Due to the rapid nature of the failure, the soil mass behaved under undrained conditions; therefore, excess pore water pressures did not have sufficient time to dissipate. Consequently, a total-stress analysis using undrained shear strength parameters is appropriate for reproducing this case.

For Chicago glacial clays, the Poisson's ratio is commonly taken as $\nu = 0.2$. Typical drained Young's modulus values range from 10~MPa to 40~MPa for medium and stiff clays, respectively \citep{Calvello2002, Kim2019}. The dilation angle was set to zero, corresponding to no volume change during yielding. The mechanical properties of the sand and clay layers are provided in Table~\ref{tab:congress_material_properties}. The surface sand layer is modeled as a drained material, while the underlying clay layers are treated as undrained. The groundwater level is defined from borehole data at a depth of 3.3~m.

\begin{figure}[H]
  \centering
  \includegraphics[width=0.8\textwidth]{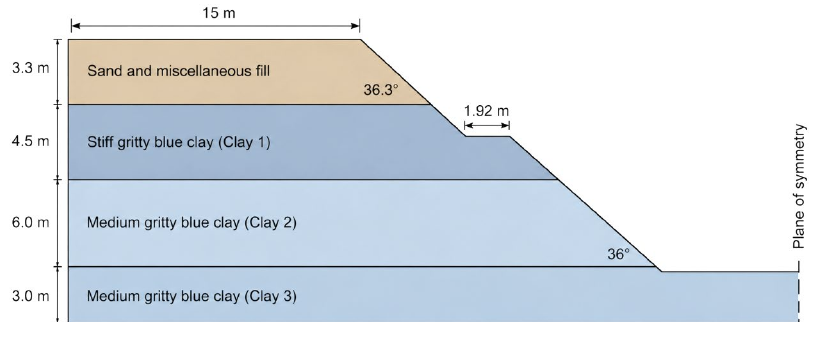}
  \caption{Geometry and geological profile of Congress Street cut.}
  \label{fig:congress_geometry}
\end{figure}

The boundary conditions include fixed constraints at the base, representing the stiff underlying stratum, and roller supports along the vertical boundaries, allowing vertical movement while restricting horizontal displacement. This case study is particularly useful because detailed geological, geometrical, and geotechnical information is available, together with clear documentation of the actual failure. The approximate dimensions of the cut at the time of failure are shown in Figure~\ref{fig:congress_geometry}. A finite element analysis is carried out using CODE\_BRIGHT \citep{Olivella1996} to investigate the influence of temperature cycles on the performance of the Congress Street cut, based on the proposed interface joint approach and the shear strength reduction technique.

\begin{table}[H]
\centering
\caption{Undrained strength parameters for the Congress Street cut \citep{Jiang2016, Li2016, Li2019}.}
\label{tab:congress_material_properties}
\renewcommand{\arraystretch}{1.15}
\small
\begin{tabular}{lcccccc}
\toprule
\textbf{Layer} 
& $\boldsymbol{\gamma}$ \textbf{[kN/m$^3$]} 
& $\boldsymbol{c}$ \textbf{[kPa]} 
& $\boldsymbol{\phi}$ \textbf{[$^\circ$]} 
& $\boldsymbol{\psi}$ \textbf{[$^\circ$]} 
& $\boldsymbol{E}$ \textbf{[kPa]} 
& $\boldsymbol{\nu}$ \\
\midrule
Sand   & 21.0 & 0  & 30 & 0 & $10^4$ & 0.3 \\
Clay 1 & 19.5 & 55 & 5  & 0 & $10^4$ & 0.3 \\
Clay 2 & 19.5 & 43 & 7  & 0 & $10^4$ & 0.3 \\
Clay 3 & 20.0 & 56 & 15 & 0 & $10^4$ & 0.3 \\
\bottomrule
\end{tabular}
\end{table}

\subsection{Numerical simulation}
\label{subsec:numerical_simulation}

\subsubsection{Viscoplastic constitutive model}
\label{subsubsec:viscoplastic_constitutive_model}

The mechanical response of the slope materials was simulated using a viscoplastic constitutive model developed for unsaturated soils. The formulation is based on the overstress viscoplasticity theory proposed by \citet{Perzyna1966} and later extended to unsaturated porous media by \citet{Desai1987}. The model captures key features relevant to partially saturated geomaterials, including time-dependent deformation, suction effects, and stress-path dependency. Therefore, it is suitable for slope stability analysis under unsaturated conditions. Table~\ref{tab:constant_viscoplastic_parameters} summarizes the parameters assumed constant for all materials, including sand and Clay 1--3, in the Congress Street cut. The viscoplastic strain rate is expressed as

\begin{equation}
\dot{\boldsymbol{\varepsilon}}^{vp}
=
\Gamma \left\langle \phi(F) \right\rangle
\frac{\partial G}{\partial \boldsymbol{\sigma}'}
\end{equation}

where $\Gamma$ is the viscosity parameter controlling the rate dependency, $\phi(F)$ is the overstress function, $F$ is the yield function, $G$ is the viscoplastic potential, and $\boldsymbol{\sigma}'$ is the effective stress tensor. The Macaulay brackets $\langle \cdot \rangle$ ensure that viscoplastic flow occurs only when $F>0$. The overstress function is commonly written as

\begin{equation}
\phi(F)
=
\left(
\frac{F}{F_0}
\right)^N
\end{equation}
where $F_0$ is a reference stress and $N$ is a material parameter controlling the rate sensitivity. The yield function incorporates stress invariants and suction effects as

\begin{equation}
F(J_1,J_{2D},J_{3D},s)
=
a J_{2D}
-
\mu^2 F_b F_s
\end{equation}
where $J_1$, $J_{2D}$, and $J_{3D}$ are stress invariants, $a$ and $\mu$ are material parameters, $F_b$ governs the shear strength contribution, and $F_s$ accounts for suction effects. The suction-dependent term is defined as

\begin{equation}
F_s
=
\left[
1
-
\beta_s
\frac{\sqrt{27}}{2}
J_{3D}
\left(
J_{2D}^{-3/2}
\right)
\right]^m
=
(1-\beta_s s)^m
\end{equation}
where $s$ is suction, and $\beta_s$ and $m$ are material parameters. The viscoplastic potential function has a similar structure to the yield function, but includes a non-associativity parameter:

\begin{equation}
G(J_1,J_{2D},J_{3D},s)
=
a J_{2D}
-
b \mu^2 F_b F_s
\end{equation}
where $b$ controls the degree of non-associativity of the flow rule. The model is formulated in terms of net stress and suction. The hardening behavior is linked to the viscoplastic volumetric strain and is consistent with the Barcelona Basic Model (BBM):

\begin{equation}
J_1^{0}(s)
=
3p^c
\left(
\frac{J_1^{0*}}{3p^c}
\right)^{
\frac{\lambda(0)-\kappa}{\lambda(s)-\kappa}
}
\end{equation}
where $J_1^{0*}$ is the preconsolidation stress expressed in terms of the first stress invariant, $e$ is the void ratio, and $\lambda(0)$ and $\kappa$ are compressibility parameters. The hardening law depends on the viscoplastic volumetric strain as

\begin{equation}
\mathrm{d}J_1^{0*}
=
\frac{1+e}{\lambda(0)-\kappa}
J_1^{0*}
\mathrm{d}\varepsilon_v^{vp}
\end{equation}

The suction-dependent preconsolidation pressure is expressed as

\begin{equation}
p_0(s)
=
\frac{J_1^{0}(s)}{3}
\end{equation}
and depends on suction through

\begin{equation}
\lambda(s)
=
\lambda(0)
\left[
(1-r)\exp(-\beta s)+r
\right]
\end{equation}

Under the specific parameter assumptions $k_1=3k$, $k_2=3k$, $k_3=0$, and $k_4=3p_{s0}$, where $p_{s0}$ is the tensile strength of the soil under saturated conditions $(s=0)$, the yield function can be expressed in terms of mean stress $p$, deviatoric stress $q$, and suction $s$ as

\begin{equation}
F(q,p,s)
=
\frac{a}{3}q^2
-
\mu^2 \gamma^3
\left[
-
\left(
p_0(s)+ks+p_{s0}
\right)^{2-n}
\left(
p+ks+p_{s0}
\right)^n
+
\left(
p+ks+p_{s0}
\right)^2
\right]
\end{equation}
where $k$ accounts for the increase in tensile strength due to suction, and $p_{s0}$ is the tensile strength under saturated conditions. This formulation highlights the coupling between mechanical loading and suction. The shear strength parameter is also assumed to vary with suction, indicating increased strength under partially saturated conditions:

\begin{equation}
\mu(s)
=
\mu_{\mathrm{dry}}
-
\left(
\mu_{\mathrm{dry}}-\mu_{\mathrm{sat}}
\right)
\left(
\frac{\mu_{\mathrm{sat}}}{\mu_{\mathrm{dry}}}
\right)^s
\end{equation}
where $\mu_{\mathrm{dry}}$ is the critical-state line slope for dry or low-suction states, and $\mu_{\mathrm{sat}}$ is the critical-state line slope for fully saturated states. The viscoplastic model does not include cohesion $c$ explicitly. However, an equivalent apparent cohesion can be approximated by matching the yield envelope at a representative stress level, leading to

\begin{equation}
\left(J_1^{0*}\right)_F
=
\frac{6c}{\mu}
\end{equation}

This apparent cohesion arises from the suction-dependent expansion of the yield surface governed by the preconsolidation pressure.

\begin{table}[H]
\centering
\caption{Constant viscoplastic model parameters for sand and Clay 1--3.}
\label{tab:constant_viscoplastic_parameters}
\renewcommand{\arraystretch}{1.25}
\resizebox{\textwidth}{!}{%
\begin{tabular}{p{3.0cm} p{1.8cm} p{2.2cm} p{8.5cm} p{1.5cm}}
\toprule
\textbf{Category} & \textbf{Symbol} & \textbf{Units} & \textbf{Technical description} & \textbf{Value} \\
\midrule

\multirow{4}{*}{Flow and rate}
& $\Gamma_0$ & $\mathrm{s}^{-1}\mathrm{MPa}^{-1}$ 
& Fluidity: rate of viscoplastic strain. High values simulate nearly rate-independent plasticity.
& 10 \\

& $N$ & -- 
& Stress power: sensitivity of strain rate to the stress-to-yield ratio.
& 3 \\

& $F_0$ & $\mathrm{MPa}^2$ 
& Reference stress used as a normalization factor.
& 1 \\

& $b$ & -- 
& Non-associativity parameter controlling the flow direction relative to the yield surface.
& 0.1 \\

\midrule

\multirow{3}{*}{Yield surface}
& $n$ & -- 
& Shape parameter controlling the curvature of the yield locus.
& 5.1 \\

& $\gamma$ & -- 
& Scaling parameter controlling the magnitude of the yield surface.
& 0.111 \\

& $a$ & -- 
& Geometrical parameter defining the elliptic aspect of the yield locus.
& 3 \\

\midrule

\multirow{5}{*}{\makecell{Hardening and\\ softening}}
& $\kappa$ & -- 
& Elastic stiffness parameter corresponding to the slope of the swelling/recompression line.
& 0.01 \\

& $\lambda(0)$ & -- 
& Plastic stiffness parameter corresponding to the slope of the virgin compression line at zero suction.
& 0.1 \\

& $r$ & -- 
& Suction limit parameter defining the stiffness limit at high suction.
& 0.7 \\

& $\beta$ & -- 
& Hardening parameter controlling the rate of stiffness variation with suction.
& 5 \\

& $p^c$ & MPa 
& Reference pressure defining the stress level of the LC curve.
& 0.84 \\

\bottomrule
\end{tabular}%
}
\end{table}

Table~\ref{tab:calibrated_viscoplastic_parameters} lists the parameters adopted for each layer after the back-analysis procedure. These parameters were calibrated to ensure that the numerical model reproduces the observed field behavior with acceptable accuracy.

\begin{table}[H]
\centering
\caption{Calibrated viscoplastic model parameters for sand and Clay 1--3.}
\label{tab:calibrated_viscoplastic_parameters}
\renewcommand{\arraystretch}{1.15}
\small
\begin{tabular}{lcccccc}
\toprule
\textbf{Type} 
& $\boldsymbol{\mu_{\mathrm{dry}}}$ 
& $\boldsymbol{J_1^{0*}}$ \textbf{[MPa]} 
& $\boldsymbol{\mu_{\mathrm{sat}}}$ 
& $\boldsymbol{k_1}$ 
& $\boldsymbol{k_2}$ 
& $\boldsymbol{k_4}$ \\
\midrule
Sand  & 1.55 & 0.01 & 1.19 & 0.3 & 0.3 & 0.03 \\
Clay 1 & 0.45 & 0.84 & 0.17 & 0.3 & 0.3 & 0.03 \\
Clay 2 & 0.63 & 0.45 & 0.25 & 0.3 & 0.3 & 0.03 \\
Clay 3 & 1.01 & 0.27 & 0.56 & 0.3 & 0.3 & 0.03 \\
\bottomrule
\end{tabular}

\vspace{0.4em}
\begin{minipage}{0.95\textwidth}
\footnotesize
\textit{Note:} $\mu_{\mathrm{dry}}$ is the critical-state line slope in dry conditions; 
$J_1^{0*}$ is the initial intercept of the yield surface on the $J_1$ axis, equal to $p_0^*$ in BBM; 
$\mu_{\mathrm{sat}}$ is the critical-state line slope for fully saturated conditions; 
$k_1$ accounts for the increase in tensile strength due to suction and is usually equal to $3k$; 
$k_2$ ensures that the plastic potential evolves at the same rate as the yield surface and is usually equal to $3k$; 
$k_4$ is the initial tensile strength and is usually equal to $3p_{s0}$. 
Since CODE\_BRIGHT uses the first stress invariant $J_1$, and $J_1=3p$, any parameter defined in $p$-space must be multiplied by 3 to be used in the $J_1$ formulation. Therefore, $J_{1,\mathrm{tensile}}=3p_{s0}+3ks$.
\end{minipage}
\end{table}

\subsubsection{Strength reduction method}
\label{subsubsec:strength_reduction_method}

The stability of the slope is evaluated using the strength reduction method (SRM). Unlike traditional Mohr--Coulomb limit-equilibrium analyses, which typically reduce the shear strength parameters $c'$ and $\phi'$ until the slope fails, the SRM implemented here uses the viscoplastic framework to identify the factor of safety (FoS).

In this study, a generalized reduction approach is adopted. Rather than reducing only the shear parameters, the entire yield surface is scaled. In the viscoplastic model, strength is governed by the parameters $\mu$, $p_s$, and $\left(J_1^{0*}\right)_F$. This approach allows a coupled assessment of both shear strength and volumetric yield limits. Therefore, both the frictional--cohesive strength and the preconsolidation pressure of the clay are included in the stability analysis. The factored parameters are defined as

\begin{equation}
\begin{aligned}
\mu_f
&=
\frac{\mu}{\mathrm{FoS}}
=
\frac{1}{\mathrm{FoS}}
\left(
\frac{6\sin\phi'}{3-\sin\phi'}
\right)
\\[0.5em]
p_{s,f}
&=
\frac{p_s(s)}{\mathrm{FoS}}
=
\frac{p_{s0}+ks}{\mathrm{FoS}}
\\[0.5em]
\left(J_1^{0*}\right)_{F,f}
&=
\frac{\left(J_1^{0*}\right)_F}{\mathrm{FoS}} 
\end{aligned}
\label{eq:srm_reduced_parameters}
\end{equation}
where $\mu_f$ is the reduced slope of the yield surface in the stress-invariant space, $p_{s,f}$ is the reduced tensile limit, incorporating both the saturated tensile strength $p_{s0}$ and the suction-induced contribution $ks$. In CODE\_BRIGHT, this corresponds to reducing the input parameters $k_1$, $k_2$, and $k_4$. The parameter $\left(J_1^{0*}\right)_{F,f}$ is the reduced preconsolidation pressure, representing the shrinkage of the compression cap.

Reducing $\left(J_1^{0*}\right)_F$ is particularly important for Clay 1--3 because it prevents the model from overestimating strength in the deeper, high-pressure zones of the slope, where the material may lose a significant part of its preconsolidation pressure once movement initiates. The FoS is determined by incrementally increasing the reduction factor until an asymptotic increase in the nodal displacement at the slope crest is observed.

\begin{figure}[H]
  \centering
  \includegraphics[width=0.6\textwidth]{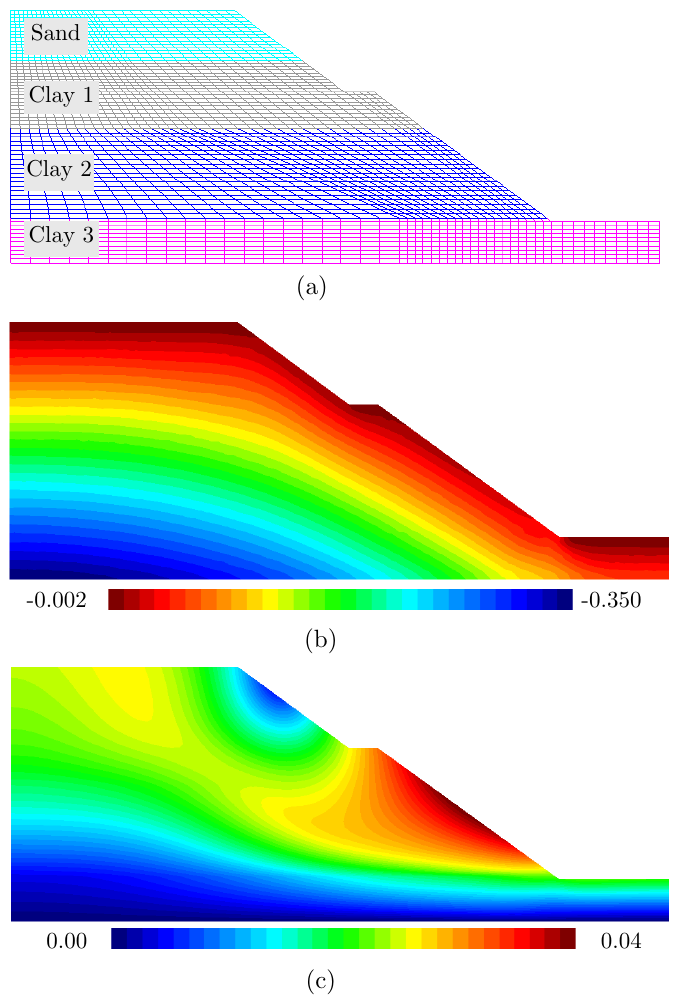}
  \caption{(a) The mesh distribution of the model and defined related joints with zero thickness elemenst, (b) the initial contours of the vertical stress (MPa), and (c) the slope deformation (m) in initial condition.}
  \label{fig:initial_model_results}
\end{figure}

Figure~\ref{fig:initial_model_results} illustrates the discretized domain, where mesh refinement is applied near the slope surface to improve accuracy. The mesh consists of 2,574 quadrilateral elements and 2,766 nodes. Initial conditions are established by prescribing a linear distribution of pore water pressure and in-situ stresses from the crest to the base of the slope. The model is first run over a 20-day initialization period to reach equilibrium, after which all displacements are reset to zero.

Figure~\ref{fig:initial_model_results}b presents the vertical stress distribution at the equilibrium state, before applying thermal loading. At this stage, the slope shows a maximum displacement of approximately 4~cm, located within the middle of the second clay layer, Clay~2. Figure~\ref{fig:initial_model_results}c shows the corresponding deformation pattern under these initial conditions, representing the stabilized configuration before the slope stability analysis.

The analysis indicates that the slope is close to failure, with a computed factor of safety of approximately 1.35, reflecting marginal stability. This result is consistent with values reported using Bishop's limit-equilibrium method \citep{Jiang2016, Li2016} and finite-element strength-reduction approaches in previous studies \citep{Hammah2005, Wang2021}.

\begin{figure}[H]
  \centering
  \begin{subfigure}[t]{0.6\textwidth}
    \centering
    \includegraphics[width=\textwidth]{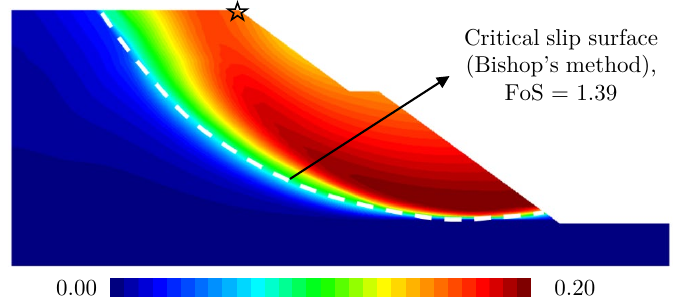}
    \caption{}
    \label{fig:srm_results_a}
  \end{subfigure}
  \begin{subfigure}[t]{0.55\textwidth}
    \centering
    \includegraphics[width=\textwidth]{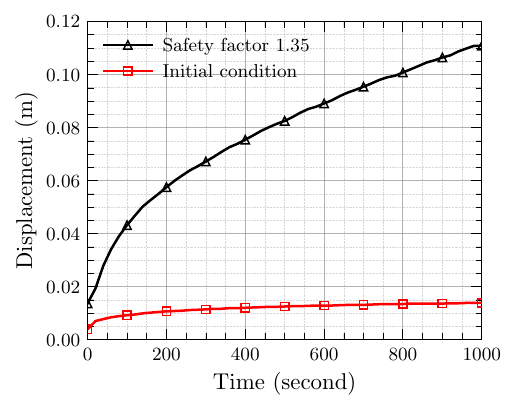}
    \caption{}
    \label{fig:srm_results_b}
  \end{subfigure}

  \caption{Displacement contours and critical slip surface obtained from the shear strength reduction analysis using CODE\_BRIGHT: 
  \textbf{(a)} comparison between the CODE\_BRIGHT result, with $\mathrm{FoS}=1.35$, and the slip surface derived from Bishop's method, with $\mathrm{FoS}=1.39$; 
  \textbf{(b)} time evolution of displacement at a representative point.}
  \label{fig:srm_results}
\end{figure}

The displacement contours in Figure~\ref{fig:srm_results}a show that the failure mechanism obtained from the strength-reduction analysis is not purely circular, but rather forms a compound surface. It consists of a semi-circular segment passing through the sand layer and the upper clay strata, followed by a more planar sliding component developing along the interface between the second and third clay layers, particularly toward the slope toe. The deformation pattern shows a well-defined failure mechanism, with maximum displacement concentrated along a continuous band extending from the upper part of the slope toward the toe.

In contrast, the slip surface predicted by Bishop's method, shown as a dashed line in Figure~\ref{fig:srm_results}a, remains fully circular and does not capture the planar sliding component observed in the finite-element results. Despite these differences in the predicted geometry of the failure surface, the overall agreement in terms of the factor of safety and the general failure pattern supports the reliability of the adopted model parameters. This consistency suggests that the numerical model can realistically reproduce the slope behavior and provides a solid basis for further analysis.

Figure~\ref{fig:srm_results}b illustrates the time evolution of displacement at a representative point located at the crest of the Congress Street cut. Under the initial condition, displacements remain very small and stabilize quickly, indicating that the slope is initially in equilibrium. In contrast, when the strength-reduction technique is applied, corresponding to $\mathrm{FoS}=1.35$, the displacement increases significantly with time and shows a nonlinear growth trend. The rapid increase at early times, followed by a gradual reduction in the deformation rate, indicates the development of plastic strains along the failure surface. This behavior is consistent with a progressive failure mechanism, in which deformations accumulate until the system approaches instability.

\subsubsection{Zero-thickness elements}
\label{subsubsec:zero_thickness_elements}

\begin{figure}[H]
  \centering
  \includegraphics[width=0.85\textwidth]{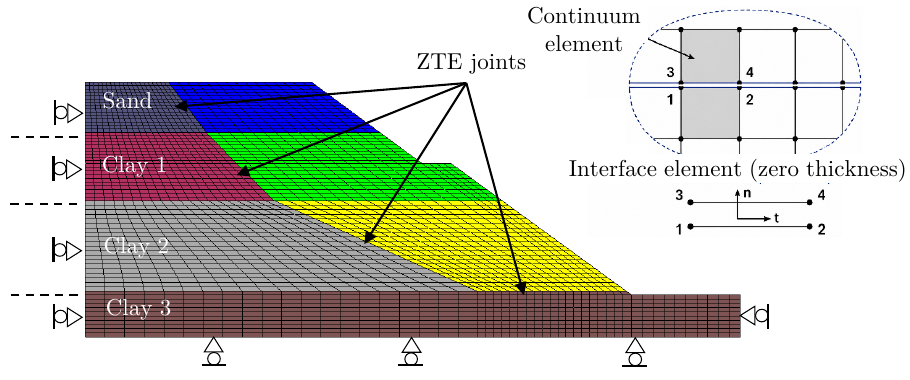}
  \caption{Mesh and soil stratigraphy of the slope, boundary conditions, and the location of zero-thickness elements (ZTE joints) inserted along the predefined critical slip surface to model potential sliding and displacement discontinuities.}
  \label{fig:zte_interface}
\end{figure}

Zero-thickness elements (ZTE joints) are introduced along the potential failure surface previously identified from the safety factor analysis, as illustrated in Figure~\ref{fig:zte_interface}. These elements do not have physical thickness and allow displacement discontinuities to develop while transmitting stresses across the interface. The main role of the ZTE formulation is to explicitly simulate the initiation and evolution of sliding.

Unlike standard continuum elements, ZTE joints can capture relative shear displacement between adjacent soil blocks. Therefore, they are particularly suitable for representing strain localization and progressive failure. They also allow specific constitutive laws to be assigned to the interface, including reduced shear strength or softening behavior, which governs the onset of slip. In this study, these interface elements are used to explicitly represent the discontinuity where shear localization and sliding are most likely to occur. Their placement follows the predicted slip path, allowing the model to capture the initiation and evolution of displacement along this surface.

The ZTE formulation enables a more realistic simulation of progressive failure mechanisms by accounting for relative movements and possible strength degradation along the interface. This approach also makes it possible to track the development of shear stresses and displacements under subsequent loading conditions, including the influence of thermal effects on slope response.

Table~\ref{tab:constant_zte_parameters} details the fixed mechanical and elastic properties assigned to the ZTE joints representing the sand and clay layers, Clay 1--3. These constitutive parameters remain constant for their respective layers throughout the simulation. In contrast, Table~\ref{tab:calibrated_zte_parameters} summarizes the layer-specific parameters obtained through back-analysis. The calibration was performed to ensure numerical consistency between the jointed model and the baseline reference model without ZTE joints.

\begin{table}[H]
\centering
\caption{Constant ZTE joint model parameters for sand and Clay 1--3.}
\label{tab:constant_zte_parameters}
\renewcommand{\arraystretch}{1.20}
\small
\resizebox{\textwidth}{!}{%
\begin{tabular}{p{1.6cm} p{2.0cm} p{9.5cm} p{1.6cm}}
\toprule
\textbf{Symbol} & \textbf{Units} & \textbf{Description} & \textbf{Value} \\
\midrule
$u_c^*$ 
& m 
& Critical shear displacement for cohesion, at which cohesion begins to degrade or vanish. 
& 0.015 \\

$u_{\phi}^*$ 
& m 
& Critical shear displacement for friction, defining the threshold for transition from peak to residual friction. 
& 0.015 \\

$q_u$ 
& MPa 
& Compression strength at which dilatancy vanishes. 
& 151 \\

$\Gamma_0$ 
& $\mathrm{s}^{-1}$ 
& Fluidity parameter governing the viscous flow or creep of the interface. 
& $1\times10^{-4}$ \\

$N$ 
& -- 
& Power of the stress function, controlling the nonlinearity of the stress--strain relationship. 
& 2 \\

$m$ 
& MPa 
& Model parameter used as a base stiffness modulus for calculating the elastic response of the joint. 
& 1 \\

$K_s$ 
& $\mathrm{MPa}\,\mathrm{m}^{-1}$ 
& Shear stiffness, defined as the ratio of shear stress to relative shear displacement of the joint. 
& 10 \\

$E$ 
& $\mathrm{MPa}\,\mathrm{m}^{-1}$ 
& Out-of-plane stiffness governing deformation perpendicular to the primary shearing direction. 
& 2500 \\

$a_{\min}$ 
& m 
& Minimum aperture, representing the minimum possible gap between joint surfaces. 
& 0.00009 \\

$a_0$ 
& m 
& Initial aperture of the joint, representing the initial distance between the two joint faces. 
& 0.0001 \\
\bottomrule
\end{tabular}%
}
\end{table}

\begin{table}[H]
\centering
\caption{Calibrated ZTE joint model parameters for sand and Clay 1--3.}
\label{tab:calibrated_zte_parameters}
\renewcommand{\arraystretch}{1.15}
\small
\begin{tabular}{lccccc}
\toprule
\textbf{Type} 
& $\boldsymbol{c_0}$ \textbf{[MPa]} 
& $\boldsymbol{\phi_0}$ \textbf{[$^\circ$]} 
& $\boldsymbol{\phi_{\mathrm{res}}}$ \textbf{[$^\circ$]} 
& $\boldsymbol{c_1}$ \textbf{[MPa]} 
& $\boldsymbol{t_1}$ \\
\midrule
Sand   & 0.001 & 30 & 27 & 0.1 & 0.1 \\
Clay 1 & 0.055 & 5  & 5  & 0.1 & 0.1 \\
Clay 2 & 0.043 & 7  & 7  & 0.1 & 0.1 \\
Clay 3 & 0.056 & 15 & 13 & 0.1 & 0.1 \\
\bottomrule
\end{tabular}

\vspace{0.4em}
\begin{minipage}{0.92\textwidth}
\footnotesize
\textit{Note:} $c_0$ is the initial cohesion; $\phi_0$ is the initial friction angle; 
$\phi_{\mathrm{res}}$ is the residual friction angle; $c_1$ controls the evolution of cohesion with suction; 
and $t_1$ controls the evolution of friction angle with suction.
\end{minipage}
\end{table}

To ensure consistent application of the SRM across all stress states, the strength parameters $c_0$, $\tan\phi_0$, and $\tan\phi_{\mathrm{res}}$, together with the suction-dependent parameters $t_1$ and $c_1$, are scaled by the FoS. In contrast, the stiffness parameters $K_s$ and $E$, as well as the ductility parameters $u_c^*$ and $u_{\phi}^*$, which govern the rate at which the interface transitions from peak to residual strength during shearing, are kept constant. This approach avoids numerical instability by ensuring that failure is driven by strength reduction rather than unintended changes in stiffness.

\begin{figure}[H]
  \centering

  \begin{subfigure}[t]{0.6\textwidth}
    \centering
    \includegraphics[width=\textwidth]{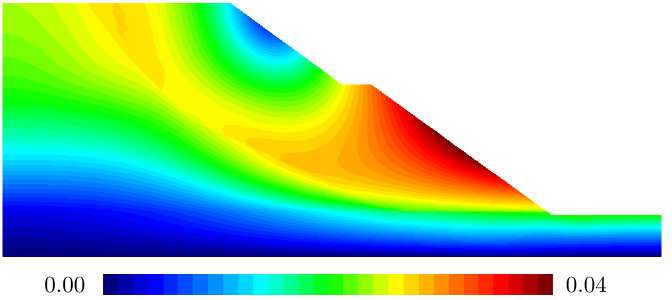}
    \caption{}
    \label{fig:zte_results_a}
  \end{subfigure}

  \vspace{0.35cm}

  \begin{subfigure}[t]{0.6\textwidth}
    \centering
    \includegraphics[width=\textwidth]{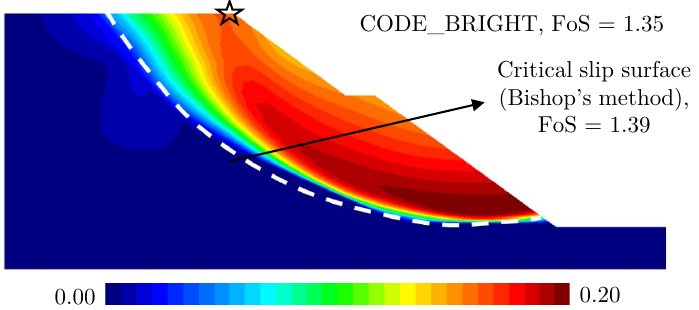}
    \caption{}
    \label{fig:zte_results_b}
  \end{subfigure}

  \vspace{0.35cm}

  \begin{subfigure}[t]{0.6\textwidth}
    \centering
    \includegraphics[width=\textwidth]{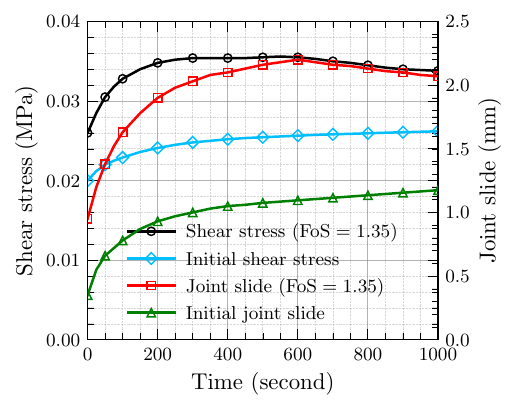}
    \caption{}
    \label{fig:zte_results_c}
  \end{subfigure}

  \caption{Slope performance with ZTE joints: 
  \textbf{(a)} initial deformation profile; 
  \textbf{(b)} comparison of displacement contours obtained from CODE\_BRIGHT, with $\mathrm{FoS}=1.35$, and the critical slip surface from Bishop's method, with $\mathrm{FoS}=1.39$; 
  \textbf{(c)} evolution of joint shear stress and sliding displacement, comparing the initial equilibrium state with the reduced-strength state at $\mathrm{FoS}=1.35$.}
  \label{fig:zte_results}
\end{figure}

Figure~\ref{fig:zte_results}a shows the corresponding deformation pattern at the initial condition, where the ZTE joints are introduced along the predefined potential failure surface. The resulting deformation pattern is broadly consistent with the case without ZTE, indicating that the overall kinematics of the slope are not artificially altered by the inclusion of interface elements. This agreement confirms that the selected mechanical parameters for the ZTE interfaces are properly calibrated and do not introduce spurious stiffness or unrealistic discontinuities into the model.

Figure~\ref{fig:zte_results}b shows the deformation contours at a strength reduction factor corresponding to $\mathrm{FoS}=1.35$, where the ZTE joints are introduced, compared with the critical slip surface derived from Bishop's method, with $\mathrm{FoS}=1.39$. The deformation pattern shows a well-defined shear band developing along a curved trajectory, closely following the predicted critical slip surface. The maximum displacement is concentrated near the toe region, which is consistent with typical rotational failure mechanisms in slopes.

The localization of deformation becomes more pronounced and sharper along the predefined interface. The shear band is more clearly aligned with the critical slip surface, indicating that the ZTE formulation effectively captures discontinuity-driven failure. Compared with Figure~\ref{fig:srm_results}a, the deformation field is less diffuse and more strongly concentrated along the interface. This highlights the role of ZTE joints in improving the representation of strain localization and slip development. Overall, the numerically predicted failure mechanism is in good agreement with Bishop's method. However, the inclusion of ZTE joints leads to a more realistic and physically consistent failure pattern by enforcing displacement discontinuities along potential failure planes.

Figure~\ref{fig:zte_results}c illustrates the temporal evolution of shear stress and joint slide, expressed as shear displacement, at the top of the Clay~1 layer along the ZTE interface. The results compare the initial condition of the slope with the response obtained using the strength reduction method at $\mathrm{FoS}=1.35$. At the initial stage of loading, both shear stress and joint displacement increase rapidly, indicating prompt stress redistribution and mobilization of shear resistance along the predefined failure surface. This increase is more pronounced in the $\mathrm{FoS}=1.35$ case, reflecting the effect of reduced shear strength on accelerating deformation and interface response.

With increasing time, the response gradually stabilizes. In the strength-reduced case, the shear stress reaches a peak value of approximately 0.035--0.036~MPa, followed by a slight reduction and subsequent stabilization. This response suggests the onset of post-peak softening and progressive failure along the interface. In contrast, the initial condition exhibits a lower shear stress magnitude, which increases gradually and approaches a steady value without any indication of softening, indicating stable behavior.

The evolution of joint slide further highlights the difference between the two conditions. For $\mathrm{FoS}=1.35$, the displacement increases significantly and exceeds 2.2~mm, indicating active slip along the failure surface. Conversely, the initial condition shows relatively limited displacement, remaining below approximately 1.1~mm, which is consistent with stable deformation and the absence of failure.

\subsection{Effect of thermal cycles on the Congress Street cut}
\label{subsec:thermal_cycles_congress_street}

\begin{figure}[H]
    \centering

    \begin{subfigure}[t]{0.49\textwidth}
        \centering
        \includegraphics[width=\textwidth]{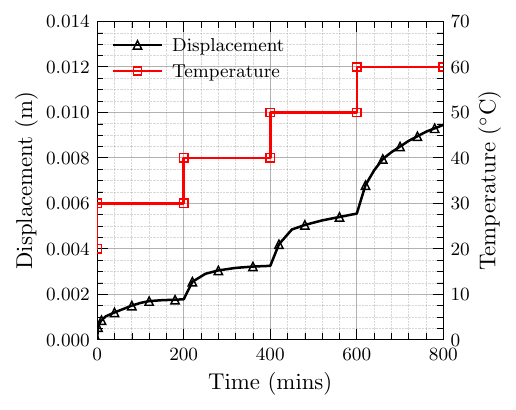}
        \caption{}
        \label{fig:thermal_joint_response_a}
    \end{subfigure}
    \hfill
    \begin{subfigure}[t]{0.49\textwidth}
        \centering
        \includegraphics[width=\textwidth]{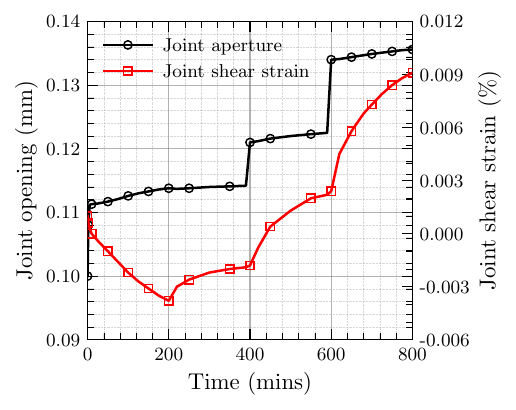}
        \caption{}
        \label{fig:thermal_joint_response_b}
    \end{subfigure}

    \vspace{0.4cm}

    \begin{subfigure}[t]{0.55\textwidth}
        \centering
        \includegraphics[width=\textwidth]{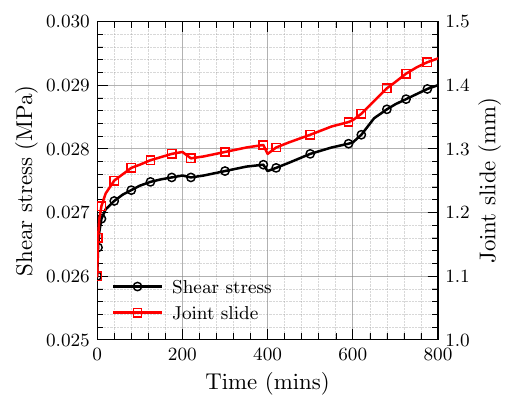}
        \caption{}
        \label{fig:thermal_joint_response_c}
    \end{subfigure}

    \caption{%
    \textbf{(a)} Evolution of displacement at the slope crest during the heating sequence,
    \textbf{(b)} evolution of joint aperture and joint shear strain with time, and
    \textbf{(c)} evolution of shear stress and joint slip.}
    \label{fig:thermal_joint_response}
\end{figure}

After verification of the model parameters, the slope response is evaluated under successive thermal cycles of 20, 30, 40, 50, and 60~$^\circ$C. Based on the proposed temperature-dependent interface resistance relationship, given in Equation~\eqref{eq:temperature_dependent_interface_resistance}, both cohesion and residual friction angle decrease with increasing temperature for each soil layer. The corresponding strength parameters for each thermal cycle are summarized in Table~\ref{tab:thermal_zte_parameters}. For each thermal cycle, a time interval of approximately 200~min \citep{Badakhshan2024, Badakhshan2026} is considered sufficient for the slope to reach a quasi-stable condition before applying the next temperature increment, as illustrated in Figure~\ref{fig:thermal_joint_response}a.

\begin{landscape}
\begin{table}[p]
\centering
\caption{ZTE joint parameters for sand and Clay 1--3 after applying the thermal cycles.}
\label{tab:thermal_zte_parameters}
\renewcommand{\arraystretch}{1.25}
\setlength{\tabcolsep}{4.2pt}
\small

\begin{threeparttable}
\begin{tabular}{lccc ccc ccc ccc ccc ccc}
\toprule
\multirow{2}{*}{\textbf{Type}}
& \multirow{2}{*}{$\boldsymbol{\mu_{fK}}$, $\boldsymbol{\mu_{fs}}$}
& \multirow{2}{*}{$\boldsymbol{\beta_d}$}
& \multirow{2}{*}{$\boldsymbol{\xi_{\phi}}$}
& \multicolumn{3}{c}{\textbf{20~$^\circ$C} $(T_{\mathrm{ref}})$}
& \multicolumn{3}{c}{\textbf{30~$^\circ$C}}
& \multicolumn{3}{c}{\textbf{40~$^\circ$C}}
& \multicolumn{3}{c}{\textbf{50~$^\circ$C}}
& \multicolumn{3}{c}{\textbf{60~$^\circ$C}} \\
\cmidrule(lr){5-7}
\cmidrule(lr){8-10}
\cmidrule(lr){11-13}
\cmidrule(lr){14-16}
\cmidrule(lr){17-19}
&
&
&
& \makecell{$c_0$\\{[MPa]}} 
& \makecell{$\phi_0$\\{[$^\circ$]}} 
& \makecell{$\phi_{\mathrm{res}}$\\{[$^\circ$]}}
& \makecell{$c_0$\\{[MPa]}} 
& \makecell{$\phi_0$\\{[$^\circ$]}} 
& \makecell{$\phi_{\mathrm{res}}$\\{[$^\circ$]}}
& \makecell{$c_0$\\{[MPa]}} 
& \makecell{$\phi_0$\\{[$^\circ$]}} 
& \makecell{$\phi_{\mathrm{res}}$\\{[$^\circ$]}}
& \makecell{$c_0$\\{[MPa]}} 
& \makecell{$\phi_0$\\{[$^\circ$]}} 
& \makecell{$\phi_{\mathrm{res}}$\\{[$^\circ$]}}
& \makecell{$c_0$\\{[MPa]}} 
& \makecell{$\phi_0$\\{[$^\circ$]}} 
& \makecell{$\phi_{\mathrm{res}}$\\{[$^\circ$]}} \\
\midrule
Sand   
& 0.5 & 100 & 0.5 
& 0.001 & 30  & 27  
& 0     & 24  & 21  
& 0     & 19  & 16  
& 0     & 16  & 13  
& 0     & 13  & 9   \\

Clay 1 
& 0.5 & 100 & 0.5 
& 0.055 & 5   & 5   
& 0.043 & 4   & 3   
& 0.035 & 3   & 2   
& 0.029 & 2.7 & 1.7 
& 0.024 & 2.2 & 1.2 \\

Clay 2 
& 0.5 & 100 & 0.5 
& 0.043 & 7   & 7   
& 0.034 & 5.5 & 4.5 
& 0.028 & 4.5 & 3   
& 0.023 & 3.7 & 1.8 
& 0.019 & 3   & 1.4 \\

Clay 3 
& 0.5 & 100 & 0.5 
& 0.056 & 15  & 13  
& 0.044 & 12  & 9   
& 0.036 & 9   & 7   
& 0.030 & 8   & 5   
& 0.025 & 6.7 & 4   \\
\bottomrule
\end{tabular}

\begin{tablenotes}
\footnotesize
\item \textit{Note:} $\mu_{fK}$ and $\mu_{fs}$ are dimensionless parameters governing the rate of stiffness degradation with temperature. 
$\beta_d$ is the degradation parameter controlling the rate of softening. 
$\xi_{\phi}$ is the non-isothermal parameter controlling the temperature dependence of the friction angle.
\end{tablenotes}

\end{threeparttable}
\end{table}
\end{landscape}

Figure~\ref{fig:thermal_joint_response}a presents the evolution of displacement at the slope crest during the heating sequence. The displacement increases progressively with temperature. At lower temperatures, between 20 and 30~$^\circ$C, the slope exhibits relatively small and gradual deformation, indicating that the reduction in shear strength remains limited. However, as the temperature increases beyond 40~$^\circ$C, a more pronounced displacement response is observed, reflecting significant degradation of the interface strength. The largest deformation occurs at 60~$^\circ$C, where the slope shows an accelerated displacement trend, indicating reduced stability and increased susceptibility to failure. The stepwise nature of the displacement curve corresponds directly to the imposed thermal cycles, with each plateau indicating temporary stabilization before the next heating stage. This behavior highlights the cumulative effect of thermal loading, where repeated exposure to elevated temperature progressively weakens the interface properties and amplifies deformation.

Figure~\ref{fig:thermal_joint_response}b illustrates the evolution of joint aperture and joint shear strain with time. The joint aperture shows a stepwise increase, closely following the imposed thermal cycles. At the beginning of the analysis, a slight opening of the joint is observed due to initial thermal adjustment and stress redistribution. As temperature increases, the aperture gradually increases, with distinct jumps occurring at higher temperature stages. After each heating stage, the joint does not return to its initial condition, indicating irreversible deformation and progressive opening. The most significant increase in aperture occurs at 60~$^\circ$C, showing that thermal effects become more pronounced at elevated temperatures.

In contrast, the joint shear strain initially shows a slight reduction during the early stage of loading. This behavior can be attributed to local adjustment and rearrangement at the interface after the first thermal exposure. However, as the thermal cycles continue, the shear strain increases steadily and then accelerates after the higher temperature stages. This trend indicates that the interface undergoes progressive degradation, transitioning from a relatively stable condition to a weakened state with increased susceptibility to shear deformation.

Figure~\ref{fig:thermal_joint_response}c shows the evolution of shear stress and joint slip. The shear stress exhibits an overall increasing trend throughout the thermal cycles, although its rate of increase is moderate compared with the deformation parameters. This behavior indicates that the interface continues to mobilize resistance, but the efficiency of stress transfer decreases as thermal damage accumulates.

Joint slip increases consistently with time and temperature. The increase is gradual at lower temperatures, but becomes significantly steeper at 60~$^\circ$C. This acceleration is consistent with the observed increase in joint aperture and shear strain. The simultaneous increase in slip and deformation parameters indicates progressive interface weakening and loss of stiffness. The accelerated slip reflects the cumulative damage induced by repeated heating, which reduces frictional resistance and facilitates displacement along the interface.

\begin{figure}
    \centering
    \includegraphics[width=0.9\linewidth]{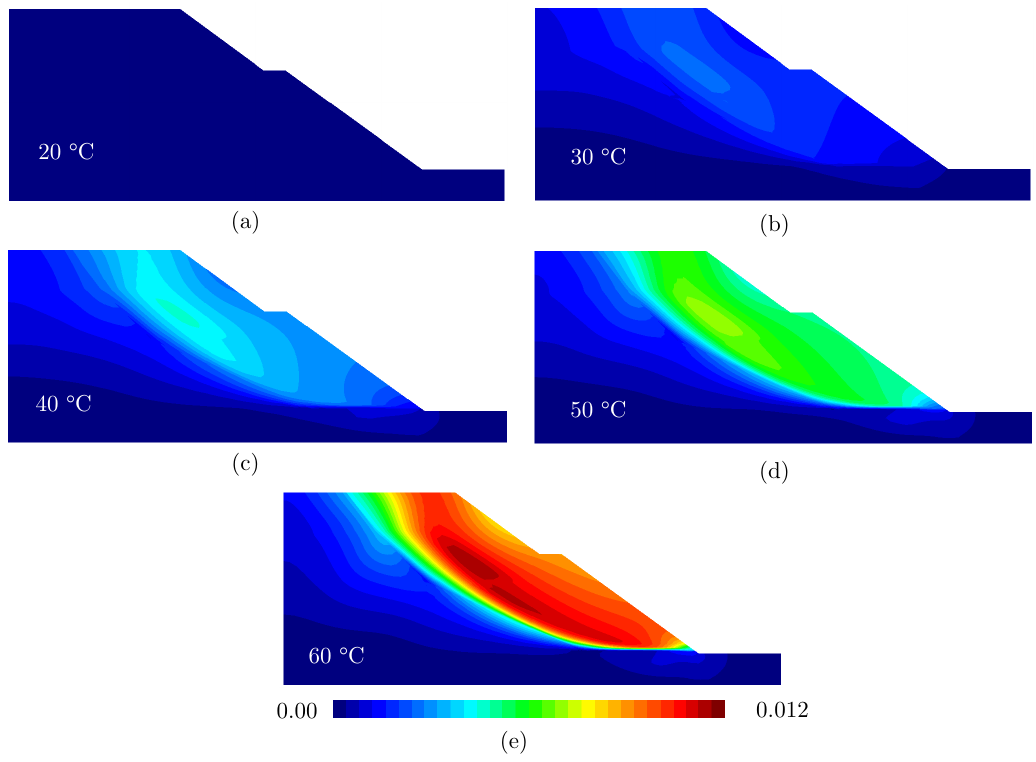}
    \caption{The deformation (m) of the slope at the end of each applied thermal cycle, a) displacement contours at 20°C, b) 30°C, c) 40°C, d) 50°C, and e) 60°C.}
    \label{fig:thermal_deformation_contours}
\end{figure}

Figure~\ref{fig:thermal_deformation_contours} illustrates the evolution of the deformation field within the slope at the end of successive thermal cycles, with temperature increasing from 20 to 60~$^\circ$C. At 20~$^\circ$C, Figure~\ref{fig:thermal_deformation_contours}a, the slope remains essentially stable, showing negligible deformation. This confirms that the initial stress state is far from failure conditions. As the temperature increases to 30~$^\circ$C, Figure~\ref{fig:thermal_deformation_contours}b, localized deformation begins to develop near the upper portion of the slope. This initial response is limited and does not yet define a clear failure mechanism.

At 40~$^\circ$C, Figure~\ref{fig:thermal_deformation_contours}c, the deformation zone expands in both magnitude and spatial extent, propagating downward along the slip surface. The contours indicate progressive strain mobilization, with higher values concentrating closer to the slope interface. A more pronounced response is observed at 50~$^\circ$C, Figure~\ref{fig:thermal_deformation_contours}d, where the deformation band becomes continuous and clearly aligns with the predefined failure path represented by the ZTE interface. This indicates that thermal loading actively drives the system toward a limit state by reducing the strength along the joint and promoting shear localization.

Finally, at 60~$^\circ$C, Figure~\ref{fig:thermal_deformation_contours}e, a well-developed and continuous deformation zone is formed, with displacement magnitudes reaching approximately 0.012~m. The concentration of deformation along the slope face and near the toe highlights the activation of a global failure mechanism. The progressive increase in deformation with temperature demonstrates that thermal effects play a critical role in weakening the slope, accelerating strain localization, and reducing overall stability. Increasing temperature not only amplifies the magnitude of deformation but also governs the transition from a stable condition to a fully developed failure mechanism.

\begin{figure}
    \centering
    \includegraphics[width=0.8\linewidth]{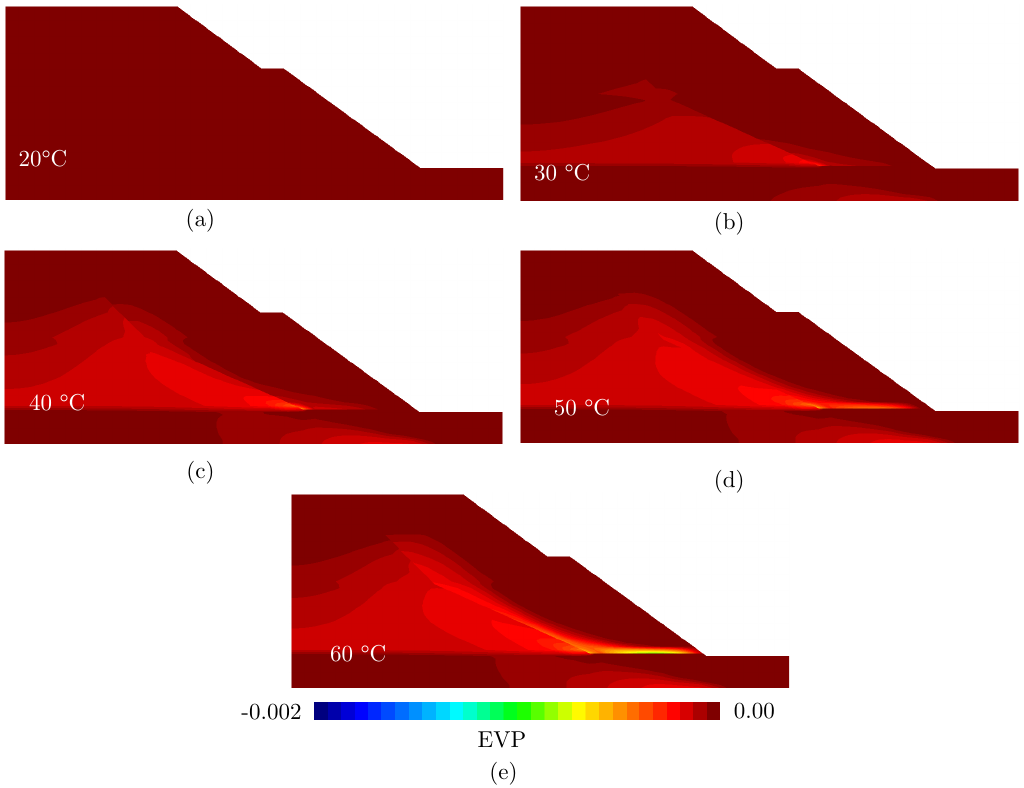}
    \caption{Evolution of plastic strain fields in the slope under increasing temperature, a) volumetric plastic strain (EVP) and b) deviatoric plastic strain (EDP) for temperatures ranging from 20 °C to 60 °C. }
    \label{fig:thermal_plastic_strains1}
\end{figure}

Figure~\ref{fig:thermal_plastic_strains1} presents the evolution of volumetric plastic strain (EVP) and deviatoric plastic strain (EDP) under increasing temperature from 20 to 60~$^\circ$C. In Figure~\ref{fig:thermal_plastic_strains1}a, the EVP contours show that volumetric plastic deformation is negligible at low temperatures, between 20 and 30~$^\circ$C, with only minor localized zones near the slope toe. As the temperature increases to 40~$^\circ$C and above, a distinct concentration of negative volumetric plastic strain develops along the potential slip surface, particularly near the toe. This localization becomes more pronounced at 50 and 60~$^\circ$C, where a continuous band of volumetric contraction is formed. This contractive behavior indicates material densification during shearing, which is consistent with the response of loose or normally consolidated soils.

\begin{figure}
    \centering
    \includegraphics[width=0.8\linewidth]{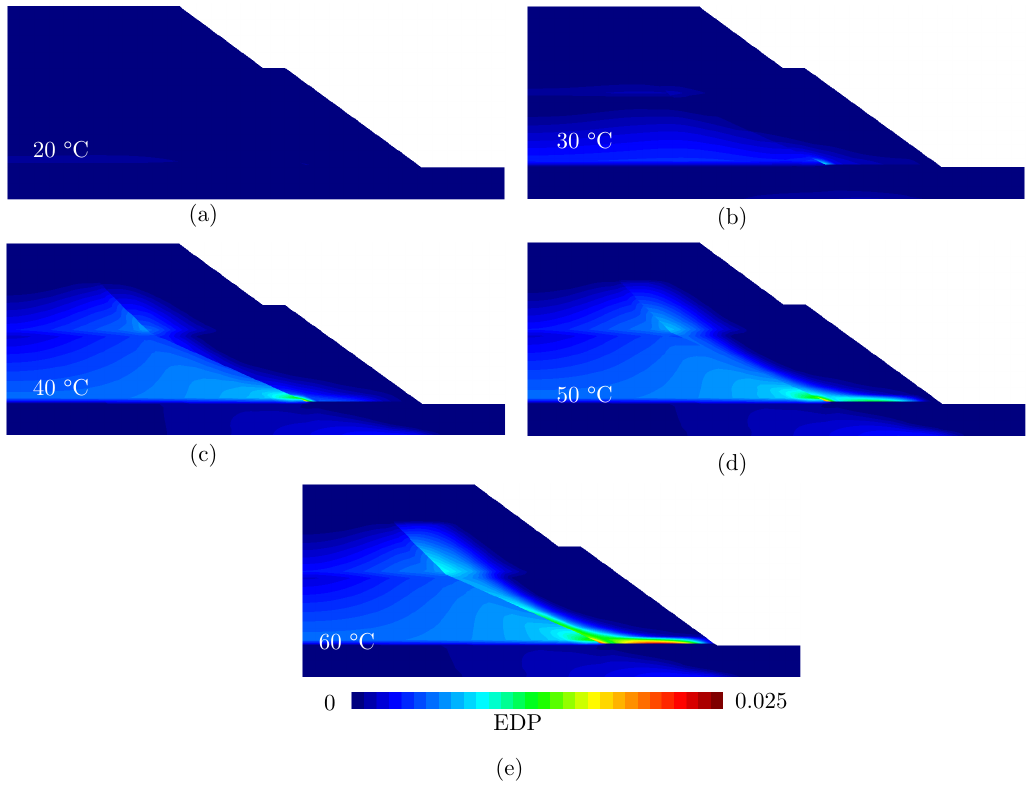}
    \caption{Evolution of plastic strain fields in the slope under increasing temperature, a) volumetric plastic strain (EVP) and b) deviatoric plastic strain (EDP) for temperatures ranging from 20 °C to 60 °C. }
    \label{fig:thermal_plastic_strains2}
\end{figure}

Figure~\ref{fig:thermal_plastic_strains2}b illustrates the evolution of EDP, representing shear-induced plastic deformation. At 20~$^\circ$C, the slope response is predominantly elastic. As the temperature increases to 30 and 40~$^\circ$C, localized zones of EDP initiate near the toe and progressively propagate upward. At higher temperatures, 50 and 60~$^\circ$C, a well-defined shear band develops, extending from the toe toward the mid-slope and closely matching a typical rotational failure mechanism. The magnitude of EDP increases significantly with temperature, reflecting progressive mobilization of shear strength and accumulation of plastic shear deformation.

A clear correlation between EVP and EDP localization is observed at elevated temperatures, confirming that the toe region acts as the initiation point of failure, with deformation propagating upward along the failure path. The slope transitions from a stable or near-elastic condition at low temperatures to a state characterized by pronounced strain localization and imminent failure at higher temperatures. Although the magnitude of EVP is notably smaller than that of EDP, indicating that shear deformation governs the failure mechanism, volumetric strain still plays a secondary but important role in the overall mechanical response.

\begin{figure}[H]
    \centering
    \includegraphics[width=0.9\linewidth]{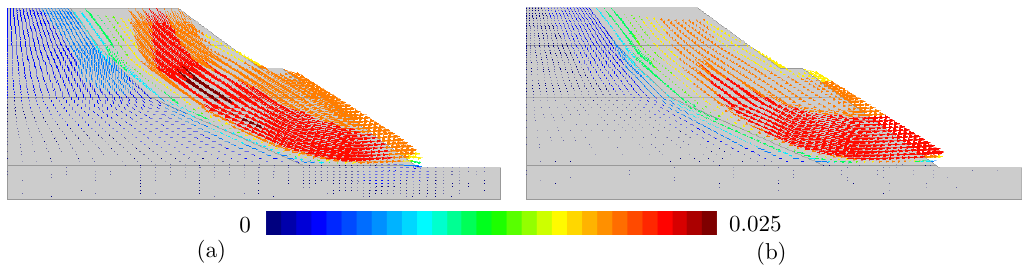}
        \vspace{0.3cm}
    \includegraphics[width=0.5\linewidth]{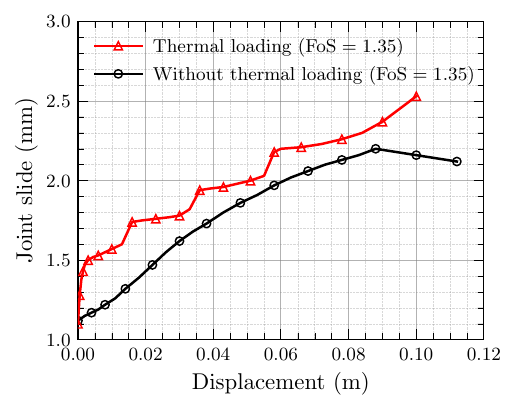}
        \\[0.01cm]
    (c)
        \caption{Displacement vector fields obtained from shear strength reduction analysis ($\mathrm{FoS}=1.35$) for \textbf{(a)} thermal loading, \textbf{(b)} without thermal loading, and \textbf{(c)} comparison of joint slide versus displacement at Clay~1 for thermal loading and a constant safety factor of 1.35.}
        \label{fig:thermal_vector_comparison}
\end{figure}

Figures~\ref{fig:thermal_vector_comparison}a and~\ref{fig:thermal_vector_comparison}b present the displacement vector fields obtained from the shear strength reduction analysis at $\mathrm{FoS}=1.35$ under thermal and non-thermal loading conditions, respectively. The comparison shows that thermal loading intensifies displacement and promotes localization of movement along the failure surface. The progressive alignment and increase in vector magnitude confirm that the slope response transitions toward a rotational failure mechanism. Moreover, the amplification of displacement near the failure surface under thermal conditions indicates a reduction in shear resistance. This results in a more pronounced and continuous failure mechanism compared with the non-thermal case. Thermal loading not only increases the magnitude of deformation, but also accelerates the development of a coherent failure surface, thereby reducing the overall stability of the slope. Figure~\ref{fig:thermal_vector_comparison}c compares the evolution of joint sliding under the strength reduction analysis at $\mathrm{FoS}=1.35$ and under thermal loading. The results indicate that temperature cycles increase the rate of joint sliding for a given level of overall slope movement. The thermally induced response shows a steeper trend, indicating more rapid mobilization of shear deformation along the joint.

As a consequence, the slope reaches higher levels of joint sliding at lower global displacements compared with the purely mechanical case. This implies that increasing temperature can reduce the stability margin of the slope. The observed response is consistent with a reduction in the apparent factor of safety under thermal effects. Therefore, neglecting thermal loading may lead to an overestimation of slope safety, particularly in environments subjected to significant temperature fluctuations.

The Congress Street cut was used as a benchmark case to investigate the influence of thermal cycles on slope performance. The adopted viscoplastic model successfully reproduced the observed behavior of the Congress Street cut, yielding a factor of safety close to previously reported values. The introduction of ZTE joints along the predefined slip surface significantly improved the simulation of strain localization and progressive failure. Compared with the continuum approach, the ZTE formulation enabled a sharper and more realistic representation of displacement discontinuities, leading to an improved prediction of the failure mechanism.

Thermal loading had a pronounced effect on slope stability. Increasing temperature caused progressive degradation of interface strength, resulting in increased displacement, joint aperture, and shear strain. This effect became particularly significant at higher temperatures, above 40~$^\circ$C, where accelerated deformation and clear strain localization were observed. The comparison between thermal and non-thermal conditions showed that thermal effects accelerated joint sliding and promoted earlier mobilization of the failure mechanism. This indicates that neglecting thermal loading may lead to an overestimation of the factor of safety.

The evolution of plastic strains confirmed that slope failure was primarily governed by shear deformation, represented by deviatoric plastic strain (EDP). Volumetric plastic strain (EVP) played a secondary but contributory role, particularly near the toe region where failure initiated. Overall, the results demonstrate that temperature-dependent interface degradation can significantly reduce slope stability and should be considered in the analysis of clayey slopes subjected to thermal fluctuations.

\section{Conclusions}
\label{sec:conclusions}

This study presented a non-isothermal viscoplastic constitutive model for clayey slip surfaces, formulated for zero-thickness interface elements. The model incorporates temperature-dependent stiffness degradation, progressive degradation of cohesion and friction angle, and rate-dependent viscoplastic slip. By linking the interface strength and stiffness to temperature, accumulated displacement, aperture evolution, and shear rate, the proposed formulation provides a consistent framework for simulating the thermo-mechanical response of clayey slip surfaces under coupled conditions.

The model was first validated against temperature-controlled drained ring-shear tests on bentonite and smectite-rich soils under heating--cooling, cooling--heating, and combined thermal paths. The simulations reproduced the main experimental trends, including thermal strengthening at slow shearing rates, thermal weakening or reduced thermal sensitivity at higher shearing rates, and the path-dependent evolution of residual shear resistance. This confirms that the proposed interface formulation can capture the combined influence of temperature and shearing rate on residual strength.

The Congress Street cut was then used as a benchmark case to assess the performance of the model in a slope stability problem. The baseline simulation reproduced the observed failure mechanism with a factor of safety close to values reported in previous studies. The introduction of zero-thickness elements along the predefined slip surface improved the representation of strain localization and progressive failure. Compared with the continuum approach, the ZTE formulation produced a sharper and more physically consistent displacement discontinuity, leading to a clearer failure mechanism.

The thermal-cycle analysis showed that increasing temperature progressively reduces the stability margin of the slope. Temperature-dependent degradation of the interface strength caused larger displacement, joint aperture, joint shear strain, and joint sliding. This effect became particularly pronounced at temperatures above 40~$^\circ$C, where accelerated deformation and clear strain localization developed along the predefined slip surface. The comparison between thermal and non-thermal simulations showed that thermal loading promotes earlier mobilization of the failure mechanism and produces higher joint sliding for a given level of global displacement.

The evolution of plastic strain fields confirmed that the failure mechanism is mainly governed by deviatoric plastic strain, while volumetric plastic strain plays a secondary but contributory role, especially near the slope toe where failure initiates. The results demonstrate that thermal loading can accelerate shear localization, reduce apparent slope stability, and lead to an overestimation of the factor of safety if ignored. Therefore, temperature-dependent interface degradation should be considered in the stability assessment of clayey slopes subjected to seasonal, climatic, geothermal, or other subsurface thermal fluctuations.

\section*{Notation}
\footnotesize
\noindent
\begin{minipage}[t]{0.48\textwidth}
\textbf{Mechanical Parameters} \\
$\delta_n$, $\delta_s$ — Normal and shear displacements \\
$\delta_n^{vp}$, $\delta_s^{vp}$ — Viscoplastic displacement components \\
$\boldsymbol{\delta}$ — Interface displacement vector \\
$\boldsymbol{\delta}_{mp}$ — Interpolated interface displacement vector \\
$\boldsymbol{{N}_{mp}^{\delta}}$ — Shape function matrix at interface \\
$\sigma'$, $\tau$ — Effective normal and shear stress \\
$c$, $c_0$ — Current and initial cohesion \\
$\phi_T$, $\phi_0$, $\phi_{res, T}$ — Peak friction angle at $T$, reference peak angle at $T_0$, and residual friction angle at $T$ \\
$\alpha$ — Yield surface parameter \\
$f_T$, $G_T$ — Yield and plastic potential functions \\
$q_u$, $\beta_d$ — Uniaxial strength and softening parameter \\
$m_v$ — Viscoplastic flow exponent \\
$\eta_v$ — Viscosity parameter \\[0.5em]

\textbf{Thermal Parameters} \\
$T$, $T_0$ — Temperature and reference temperature \\
$K_{n,T}$, $K_{s,T}$ — Temperature-dependent normal and shear stiffness \\
$K_{n,T_0}$, $K_{s,T_0}$ — Reference normal and shear stiffness \\
$\mu_{f_K}$, $\mu_{f_s}$, $\xi_\phi$ — Temperature sensitivity coefficients \\
$\lambda$, $\lambda_{\text{sat}}$, $\lambda_{\text{dry}}$ — Thermal conductivity \\
$i_c$ — Conductive heat flux \\
$j_{El}$, $j_{Eg}$ — Advective energy flux (liquid/gas) \\
$E_l$, $E_g$ — Internal energy (liquid/gas) \\
\end{minipage}
\hfill
\begin{minipage}[t]{0.48\textwidth}
\textbf{Hydraulic Parameters} \\
$k_{l}^l$, $k_{l}^t$ — Intrinsic permeability (longitudinal/transverse) \\
$k_{\text{rel}}^l$, $k_{\text{rel}}^t$ — Relative permeability (longitudinal/transverse) \\
$q_l^l$, $q_l^t$ — Advective liquid flux (longitudinal/transverse) \\
$q_g^l$, $q_g^t$ — Advective gas flux (longitudinal/transverse) \\
$s$ — Suction pressure $= P_g - P_l$ \\
$P$, $P_0$ — Air entry pressure (current/reference) \\
$\theta$, $\omega$ — Water content and mass fraction \\
$\rho$, $\mu_l$ — Fluid density and dynamic viscosity \\[0.5em]

\textbf{Interface and Coupling Parameters} \\
$r_i$ — Interface roughness coefficient \\
$\tau^*$ — Tortuosity (vapor diffusion) \\
$i_g^w$ — Non-advective vapor flux \\
$f^w$, $f^a$, $f^E$ — Source terms (water/air/energy) \\
$h_s$, $h_{s,\min}$ — Shear band thickness and minimum thickness \\
$e$ — Hydraulic aperture \\
\end{minipage}


\section*{Declaration of competing interest}
\noindent The authors declare that they have no known competing financial interests or personal relationships that could have appeared to influence the work reported in this paper.


\section*{Data availability statement}
\noindent Data will be made available on request.

\bibliographystyle{apalike}
\renewcommand{\bibname}{References}
\bibliography{References.bib}
\clearpage

%
\clearpage

\end{document}